\documentclass[%
 reprint,
 amsmath,amssymb,
 prl
]{revtex4-2}

\usepackage{float}
\usepackage{siunitx}

\usepackage{dcolumn}
\usepackage{bm}
\usepackage[utf8]{inputenc} 
\usepackage[T1]{fontenc}    
\usepackage[hidelinks]{hyperref}       
\usepackage{url}            
\usepackage{booktabs}       
\usepackage{multirow}
\usepackage{siunitx}
\usepackage{amsfonts,amsmath,amssymb}       
\usepackage{nicefrac}       
\usepackage{microtype}      
\usepackage{lipsum}		
\usepackage{graphicx}
\usepackage{natbib}
\usepackage{svg}
\usepackage{circuitikz}
\usepackage{tikz}
\usepackage{enumitem}
\tikzset{
    lb/.style = {line cap=round}
    }
\usepackage{mathtools}
\usepackage{subcaption}

\newcommand{\mrho}{\Bar{\rho}}
\newcommand{\mmu}{\Bar{\mu}}

\newcommand{\myparal}{\mathbin{\!/\mkern-5mu/\!}}

\newcommand{\rmc}{\mathrm{c}}
\newcommand{\rmd}{\mathrm{d}}
\newcommand{\rme}{\mathrm{e}}
\newcommand{\rms}{\mathrm{s}}
\newcommand{\rmt}{\mathrm{t}}
\newcommand{\rmj}{\mathrm{j}}

\newcommand{\elecCirc}[1]{
\ctikzset{bipoles/thickness=1}
    \begin{circuitikz}[line width=1.5pt]
        \draw [lb] (0,4) to[sV, -, sources/symbol/rotate=auto, l={\Large $E(t)$}] (12,4); 
        \draw (0,0) to[generic, l_={\Large $Z_{\rm c}$}] (0,4); 
        \draw [lb](12,0) to[short, -] (12,4); 
        \draw [lb](0,0) to[short, -] (3.5,0);
        \draw [lb](8.5,0) to[short, -] (12,0);
        \draw (3.5,1) to[C, l_={\Large $C_{\rm d}$}] (5.5,1);
        \draw [lb](3.5,0) to[short] (3.5,-1);
        \draw [lb](3.5,0) to[short] (3.5,1);
        \draw [lb](8.5,0) to[short] (8.5,1);
        \draw [lb](8.5,0) to[short] (8.5,-1);
        \draw (5.5,1) to[short, -] (6.5,1);
        \draw (6.5,1) to[C, l_={\Large $C_{\rm d}$}] (8.5,1);
        \draw (3.5,-1) to[R, l={\Large $R_{\rm s}$}] (8.5,-1);
        
        \draw [dashed,myred](3.0,1.5)to[short,-,l={\Large Drop}] (9.0,1.5);
        \draw [dashed,myred](3.0,-1.5)to[short,-] (9.0,-1.5);
        \draw [dashed,myred](3.0,-1.5)to[short,-] (3.0,1.5);
        \draw [dashed,myred](9.0,1.5)to[short,-] (9.0,-1.5);
        \draw [-stealth](4.5,1.8) -- (3.8,1.8);
        \node at (4.2,2.1) {\Large $\Delta\phi_1$};
        \draw [-stealth](7.5,1.8) -- (8.2,1.8);
        \node at (7.8,2.1) {\Large $\Delta\phi_2$};
    \end{circuitikz}
}

\newcommand{\UTCirc}[1]{
\ctikzset{bipoles/thickness=1}
    \begin{circuitikz}[line width=1.5pt]
        \draw (0,1) to[R,o-, l=\mbox{$R_{1} = 37.2 ~\Omega$}] (3,1);
        
        \draw (3,0) to[R,-, l=\mbox{$R_{2} = 2.31 ~\mathrm{k}\Omega$}] (6,0);
        \draw (3,2) to[C,-, l=\mbox{$C_{1} = 7.08 ~\mu \mathrm{F}$}] (6,2);

        \draw (3,0) to[short, -] (3,2);
        \draw (6,0) to[short, -] (6,2);
        \draw (6,1) to[short, -o] (7,1);    
    \end{circuitikz}
}

\definecolor{backstruct}{RGB}{240,240,240}
\definecolor{myblue}{HTML}{000779}
\definecolor{myred}{HTML}{D01122}
\definecolor{mylime}{HTML}{4FC60A}
\definecolor{mygreen}{HTML}{20b2aa}
\definecolor{mypurp}{HTML}{9932CC}
\definecolor{mywhite}{HTML}{ffffff}
\definecolor{alertcolor}{HTML}{BAA520}

\begin{document}

\preprint{APS/123-QED}

\title{Maximum Velocity of a Liquid Metal Drop Driven by a Time-Dependent Electrocapillary Excitation}

\author{Javier Otero Martinez}
\email{[jotero,agarcia]@tsc.uc3m.es} 
 \affiliation{Department of Signal Theory and Communications, Universidad Carlos III de Madrid, Spain}
 \author{Ana Garcia Armada}%
\affiliation{Department of Signal Theory and Communications, Universidad Carlos III de Madrid, Spain}

\author{Yi Li}%
\affiliation{Hybrid Materials for Opto-Electronics Group, Department of Molecules and
Materials, MESA+ Institute for Nanotechnology and Center for Brain-Inspired Nano
Systems, Faculty of Science and Technology, University of Twente, The Netherlands}
\author{Christian Nijhuis}%
\affiliation{Hybrid Materials for Opto-Electronics Group, Department of Molecules and
Materials, MESA+ Institute for Nanotechnology and Center for Brain-Inspired Nano
Systems, Faculty of Science and Technology, University of Twente, The Netherlands}
\author{Javier Rodríguez Rodríguez}\email{bubbles@ing.uc3m.es} 
\affiliation{
Department of Thermal and Fluids Engineering, ``Gregorio Millán Barbany'' Institute, Universidad Carlos III de Madrid, Spain
}

\date{\today}

\begin{abstract}
Gallium alloys are promising materials in biomedical engineering, electronics, and wireless communications, thanks to their good conductivity and their ability to sustain large deformations. They can be transported in capillaries using continuous electrowetting (CEW). Current models of CEW-driven flows do not address the transient response to fast changes in the excitation voltage, crucial in many applications. Here, we study the CEW-driven oscillatory motion of a drop of a Gallium alloy inside a capillary. We consider inertia, viscosity and the transient response of the electrical circuit consisting of the drop plus the electrolyte where it is immersed. The theory describes fairly well the experimental drop velocity and explains the existence of an optimal frequency that maximizes the velocity.
\end{abstract}

\maketitle

\textit{Introduction --} Gallium alloys exist in liquid state at room temperature. This feature, together with other physical properties, such as good electrical and thermal conductivity and low viscosity~\cite{dickeyEuteGall08, dickeyLiquMeta21}, has earned these alloys the attention of technologists and scientists. 
Another appealing characteristic of these alloys is that, when submerged in an electrolyte solution, low-voltage electric currents can significantly alter their surface tension~\cite{khanGianSwit14, eakerLiquMeta16, eakerOxidFing17}. The modification of the surface tension in response to a low-voltage (around or less than \SI{1}{\volt}) electric current, known as Continuous Electro Wetting (CEW), allows to pump the liquid metal through capillaries and microfluidic channels by purely electrical means \cite{beniContElec82}. This feature makes Gallium alloys suitable materials in applications such as liquid antennas that can dynamically change their geometry while in service \cite{goughContElec14, wangRecoLiqu15, oteromaLiquReco22}; switches for optical fiber lines \cite{jackelElecSwit83}; or, thanks to their non-toxicity, flexible and stretchable wearable medical devices \cite{zhaoSmarEute23}.



When a liquid metal drop is confined in a narrow capillary, a difference between the surface tensions of the drop's bounding menisci created by CEW induces a capillary pressure gradient inside the drop. In steady conditions, the drop then translates with a velocity that is given by a balance between the aforementioned electrocapillary pressure gradient and viscous friction. This simple principle yields a reasonably good estimation of the steady drop velocity~\cite{beniContElec82, leeSurfTens00}. However, assuming a steady motion is not a good approximation in several applications where the metal drop must respond quickly to changes in the applied voltage. This is the case, for example, of liquid antennas \cite{oteromaLiquReco22} or the generation of liquid metal droplets in microfluidic channels \cite{tangSteeLiqu15}. Beyond applications, considering the transient electrocapillarity-induced motion of a liquid metal drop is essential to answer a fundamental question that, to the best of our knowledge, has not been previously addressed: which physical effects limit the response time of a liquid metal drop to a change in the excitation voltage?

To answer the question, the simple picture where the electrocapillary-induced pressure gradient is balanced by viscous friction must be supplemented with two additional physical effects. The first one is the inertia of the two liquids involved in the motion: the liquid metal and the electrolyte solution that fills the capillary. The second effect is the transient response of the equivalent electric circuit made by the feeding electrodes, the electrolyte, and the drop. 

In this Letter we propose a simple mathematical model that takes into account these effects and thus allows us to calculate the translation velocity of a liquid metal drop in an electrolyte-filled capillary in response to a time-dependent change in the applied voltage. The model is supported by experiments where we subject the drop to sinusoidal voltage perturbations of different amplitude and frequency. In all cases, the applied voltages are kept below or equal to \SI{0.5}{\volt} in amplitude, to avoid the formation of Gallium oxide and electrolysis.

\begin{figure}[h]
    \centering
    \includegraphics[width=\columnwidth]{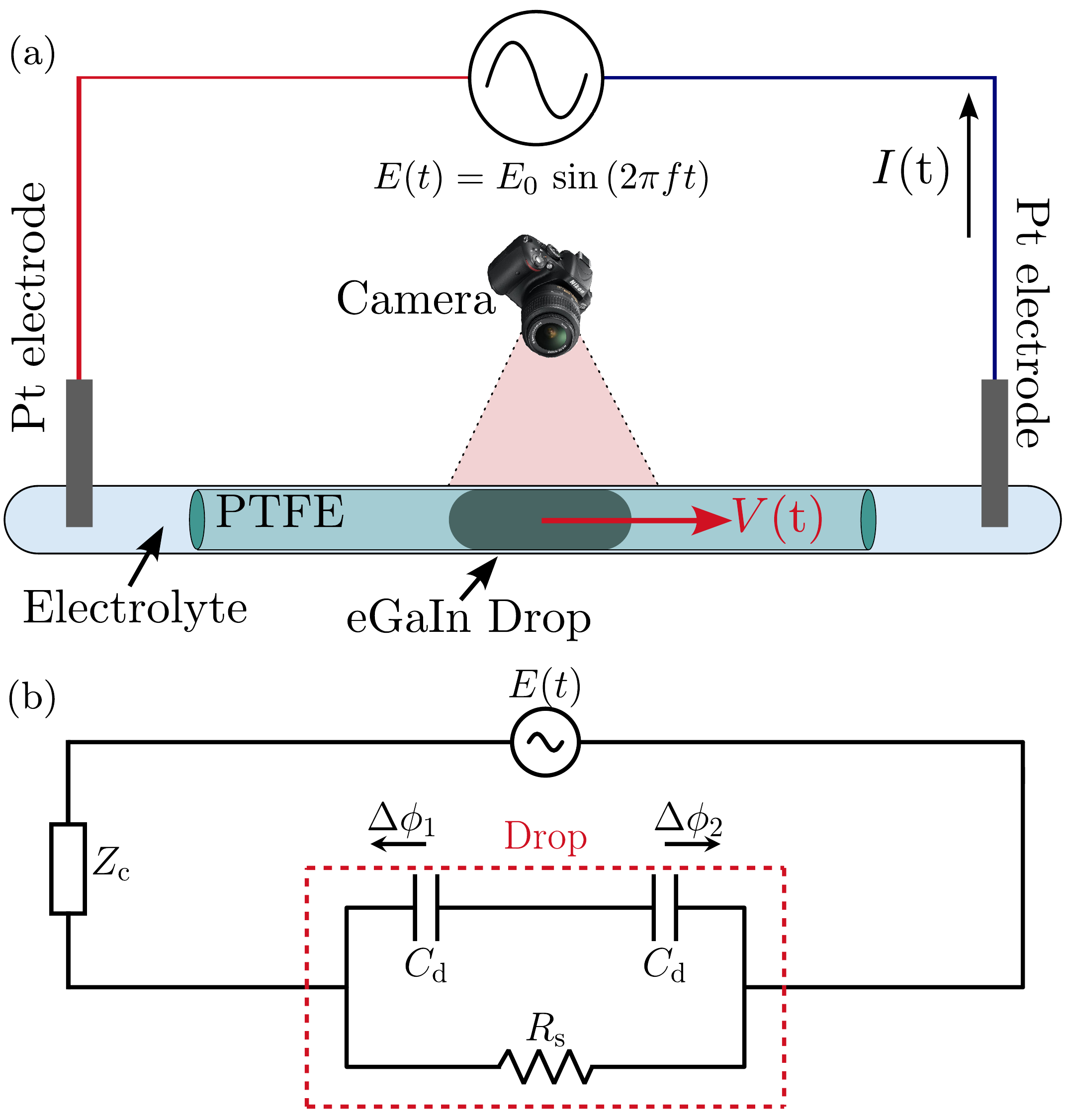}
    \caption{\label{fig:elec_cir}(a) Sketch of the experimental setup and (b) equivalent electric circuit.}
\end{figure}

\textit{Experiments --} The experimental setup (Fig.~\ref{fig:elec_cir}a) consists of a horizontal polytetrafluoroethylene (PTFE) capillary tube (length $L_t = 11$ cm, radii $R =$ \SI{0.79}{\milli\meter} and $R =$ \SI{0.4}{\milli\meter}) filled with an electrolyte, a 1 M Sodium Hydroxide (Merck Life Sciences) solution. This solution ensures electrical conductivity between the feeding electrodes and the drop, and also disolves quickly any oxide layer that may form at the surface of the liquid metal in contact with the oxygen dissolved in the water~\cite{eakerOxidFing17}. 
 By preventing the formation of a surface oxide layer, we guarantee that the rheology of the drop's surface is always Newtonian. Moreover, the properties of this electrolyte are well characterized \cite{khanGianSwit14}.
We use as liquid metal the eutectic Gallium-Indium alloy (EGaIn, 75.5\% Ga, 24\% In, Merck Life Sciences). The capillary ends connect to two small electrolyte wells (of approximately \SI{0.5}{\milli\liter}) open to the ambient. The tips of two Platinum electrodes (\SI{1}{\milli\meter} diameter Pt wire, 99.5\% purity, Merck Life Sciences) were immersed into the wells to provide the electrical excitation, $E(t) = E_0\sin\omega t$, where $E_0$ is the excitation amplitude and $\omega = 2\pi f$ its angular frequency.

The capillary and the wells equipped with the electrodes are embedded into an acrylic plate that rests on an LED screen (Edmund Optics) to provide back illumination. A video camera (Nikon D850) is placed above of the setup to record the position of the drop over time at 120 fps. After processing the acquired videos, the position, length of the drop and its instantaneous velocity are computed. The latter is obtained differentiating the measured positions with respect to the time, after smoothing the position raw data (Section S2 in \cite{suppmat}). Additionally, a calibrated resistor ($R_m=$ \SI{220}{\ohm}) in series with the electrolyte is used to measure the electrical current, $I(t)$.

\begin{figure}[H]
    \centering
    \includegraphics[width=1\linewidth]{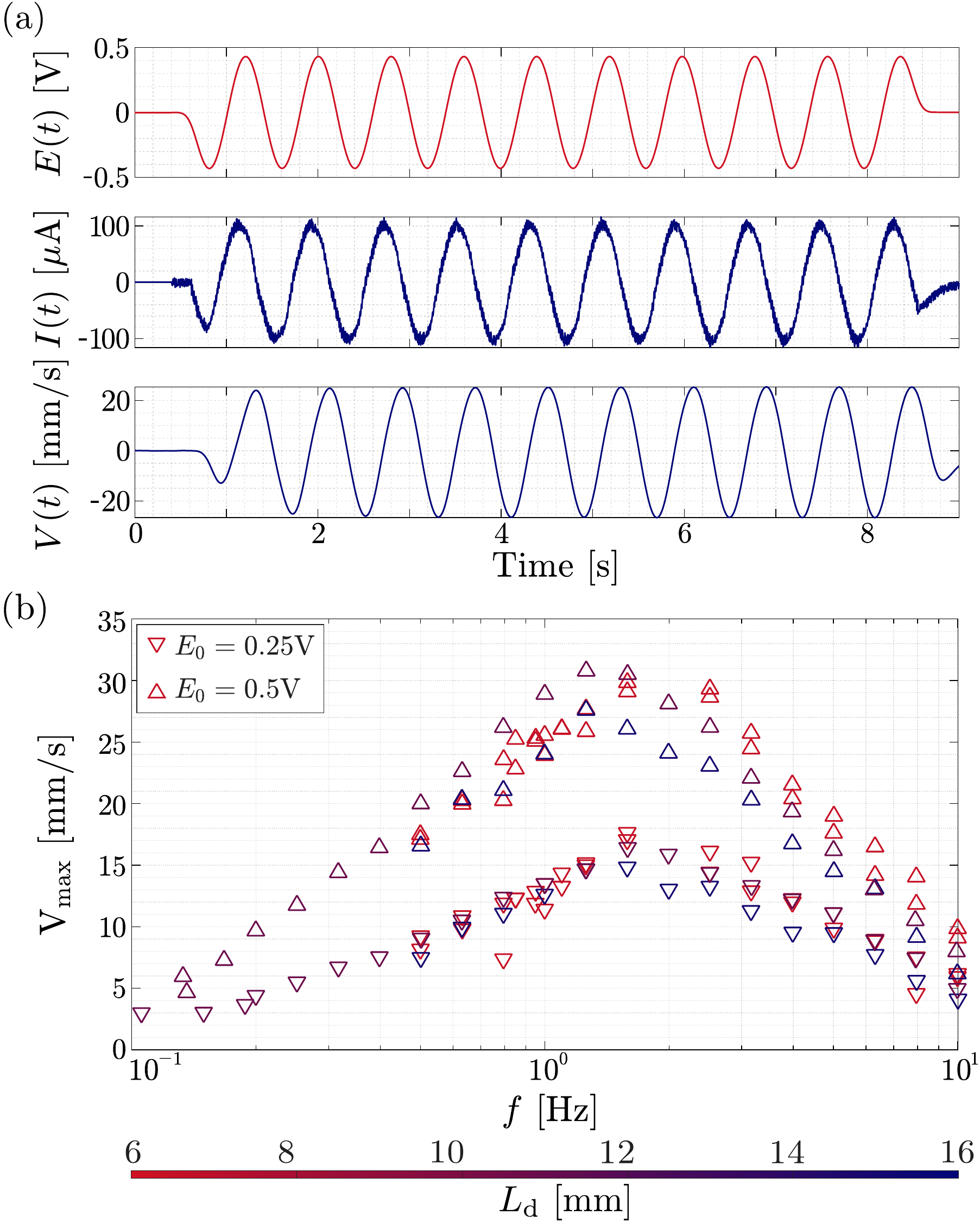} 
    \caption{\label{fig:expMeas} Experimental results corresponding to a drop of length $L_{d0} = 10.9$ mm at rest. (a) Excitation $E(t)$, measured current, $I(t)$, and velocity, $V(t)$, for an excitation of $f = 1$ Hz and amplitude $E_0=0.5$ V. (b) Peak speed $V_{\rm max}$. Colorbar represents drop length, $L_{\rmd0}$. The results correspond to the capillary with $R = $ \SI{0.79}{\milli\meter}.}
\end{figure}

During the preparation of the experiment we use a PTFE cannula connected to a syringe to gently inject the liquid metal drop (with lengths between 6 and 16 \SI{}{\milli\meter}) at the center of the capillary.  A typical experiment consists in applying a sinusoidal excitation using a signal generator and simultaneously recording both drop position and electric current. The excitation frequency, $f$, varied in the range of 0.1 to 10 Hz whereas its amplitude, $E_0$, is either \SI{0.25}{\volt} or \SI{0.5}{\volt}. We did not use higher voltages to avoid electrolysis. Indeed, we did not see any bubble formation in any of the experiments.

The outcome of a typical experiment is exemplified in Fig.~\ref{fig:expMeas}a, corresponding to a drop of length $L_{\rmd0} = 10.9$ mm in a capillary of radius $R = $ \SI{0.79}{\milli\meter} excited at $f = $ \SI{1}{\hertz} with an amplitude $E_0=0.5$ V. When a sinusoidal voltage $E(t)$ (top panel) is applied to the electrodes at the ends of the capillary, an electric current $I(t)$ (central panel) flows through the circuit. The drop responds by oscillating with a velocity $V(t)$ (bottom panel) that has the same frequency but different phase than both the excitation voltage and the current. The drop length, $L_\rmd(t)$, also exhibits a periodic oscillation, although its amplitude is very small compared to the length at rest, $L_{\mathrm{d}0}$ (typically less than 1\%, see Fig.~\ref{fig:drop_velocity_several_frequencies}b). This justifies neglecting these length variations in what follows. Both the metal and the electrolyte are incompressible, so this change in drop length reflects the thickening of the thin electrolyte layer that separates the capillary walls from the drop. We come back to the existence of this layer below when presenting the model of the electrochemical circuit (see also Section 7 in \cite{suppmat}).

Figure \ref{fig:expMeas}b shows the peak velocity $V_\mathrm{max}$, averaged over several oscillations in each experiment, as a function of the excitation frequency $f$ for drops of different lengths. For the sake of clarity, we only include in this figure the results for the $R = $ \SI{0.79}{\milli\meter} capillary. Additional results for the capillary with radius $R = $ \SI{0.4}{\milli\meter} are reported in \cite{suppmat}, section S3. The peak velocity and current (Fig.~\ref{fig:measThinCap}) are roughly proportional to the amplitude of the excitation $E_0$, while they do not show a clear dependency on the drop length. But the most remarkable feature of these results is that the peak velocity reaches a maximum value for a given frequency, a fact that to the best of our knowledge has not been reported. To explain the existence of this maximum, and to describe quantitatively the experimental results, we develop below a minimal mathematical model.

\textit{The model --} The electrical excitation creates a time-dependent capillary pressure gradient inside the drop that drives the motion, resisted by inertia and viscous stresses on both liquids. To write an expression relating the drop velocity $V$ with the excitation voltage applied on the capillary $E$, we start from the work of Quéré \cite{quereInerCapi97}, where he considers the capillary-driven invasion of a liquid into a tube in the presence of inertia and viscous forces. Applying the same equations derived in Ref. \cite{quereInerCapi97} to our two liquids and denoting by $\gamma_1$ and $\gamma_2$ the electroyte-metal interfacial tension on the left and right meniscus, respectively, we obtain
\begin{equation}
    \frac{\mathrm{d}V}{\mathrm{d}t} = \dfrac{2 \, (\gamma_1-\gamma_2)}{\mrho \, L_\rmt \, R} - V\dfrac{8\mmu}{\mrho R^2}.
    \label{eq:velocity_interfa}
\end{equation}
Here, $L_\rmt$ is the total tube length, while $\mrho = L_\rmd\rho_\rmd + \rho_\rme(L_\rmt-L_\rmd)$ and $\mmu = L_\rmd\mu_\rmd + \mu_\rme(L_\rmt-L_\rmd)$ are the effective density and viscosity of the two-liquid system, which depend on both the length of the drop and the length of the capillary tube. The subindexes "$\mathrm{e}$" and "$\mathrm{d}$" stand for electrolyte and drop (see Section 5 in \cite{suppmat} for a detailed derivation of this equation).

In the absence of any electric current, the two interfacial tensions $\gamma_1$ and $\gamma_2$ in Eq. (\ref{eq:velocity_interfa}) are identical, and the system remains at rest. When an electric potential is applied between the drop menisci, the free charges in the electrical double layers at the metal-electrolyte interfaces redistribute. This translates into a change in the energy per unit surface of the interface or, equivalently, in its interfacial tension \cite[\S25]{landauElecCont84}. This effect is commonly modeled using Lippmann's law: $\gamma_{1,2}=\gamma_0 - \,C_\rms\,(\Delta\phi_{1,2}-\phi_0)^2/2$
\cite{kirbyMicrNano10}. Here, $\gamma_0$ is the maximum interfacial tension, reached when the voltage jumps across the interface $\Delta\phi_{1,2}$ become equal to the zero-charge potential, $\Delta\phi_{1,2} = \phi_0$. Moreover, $C_\rms$ is the surface capacity of the interface. Both $C_\mathrm{s}$ and $\phi_0$ are known properties of the metal-electrolyte system \cite{khanGianSwit14}. Equation (\ref{eq:velocity_interfa}) and Lipmann's law can be combined to obtain an expression relating $V$ and $\Delta\phi_{1,2}$ (Eq.~S12, Section S5).

To close the problem we need to express the voltage jumps at the two menisci, $\Delta\phi_{1,2}$, as a function of the excitation, which requires modeling the response of the equivalent electric circuit consisting of the electrodes, the electrolyte, and the drop (Fig. \ref{fig:elec_cir}b). We denote by $Z_\rmc$ the impedance of the circuit without the drop. This impedance is determined experimentally by applying sinusoidal excitations of different frequencies and amplitudes and then, for each experiment, finding the complex impedance that best matches the measured voltage and current (Section 1 \cite{suppmat}). The electrical behavior of the drop, however, deserves more attention. At the low voltages at which we work, the two drop menisci can be modeled as pure capacitors, with a capacity $C_\rmd = 2\pi R^2 C_\rms$. Part of the total current $I(t)$ is used to vary the charge of these capacitors, but another part can flow along the drop surface, inside the thin liquid film existing between the drop and the capillary walls. This film appears when a long bubble or drop translates inside a capillary tube filled with a viscous liquid \cite{balestraViscTayl18}. In particular, its existence has been demonstrated experimentally for EGaIn drops flowing inside microfluidic devices by electrocapillarity \cite{khanInflWate14, tangSteeLiqu15}. We denote by $R_\rms$ the electrical resistance of this thin film. We neglect the resistance of the bulk metal, as it is much smaller than any other element in the circuit ($\lesssim \SI{1}{\ohm}$ using data from Ref. \cite{dickeyEuteGall08}).

The voltage difference across the drop, $\Delta\phi_2 - \Delta\phi_1$, can be readily obtained in the Fourier space using linear circuit theory on the equivalent circuit just described (Eq.~(S18), Section S5). Combining this expression with Lippmann's law and the Fourier transform of Eq.~(\ref{eq:velocity_interfa}) we get (see section S5):
\begin{equation}
    \dfrac{\Hat{V}}{V_\rmd} = \dfrac{1}{1 + \rmj\dfrac{\omega \mrho R^2}{8 \mmu}} \dfrac{1}{1 + Z_{\rm c}\left(\dfrac{1}{R_\rms} + \rmj\omega\pi R^2 C_{\rm s}\right)} \dfrac{\Hat{E}}{E_0},
    \label{eq:VelSolved}
\end{equation}

where the velocity scale 
$V_\rmd = \left|\phi_0\right| C_\rms E_0 R/4 \mmu L_\rmt.$ We can estimate this velocity scale assuming that the drop is much shorter than the capillary, thus $\mmu \approx \mu_\rme$. Using the values provided in Table \ref{tab:params}, $V_\rmd \approx$ \SI{36}{\centi\meter\per\second\per\volt} and $V_\rmd \approx$ \SI{18}{\centi\meter\per\second\per\volt} for $R =$ \SI{0.79}{\milli\meter} and $R =$ \SI{0.4}{\milli\meter} respectively, which is a reasonably good estimation of the typical velocities measured in our experiments.

\begin{figure}[H]
    \centering
    \includegraphics[width=1\linewidth]{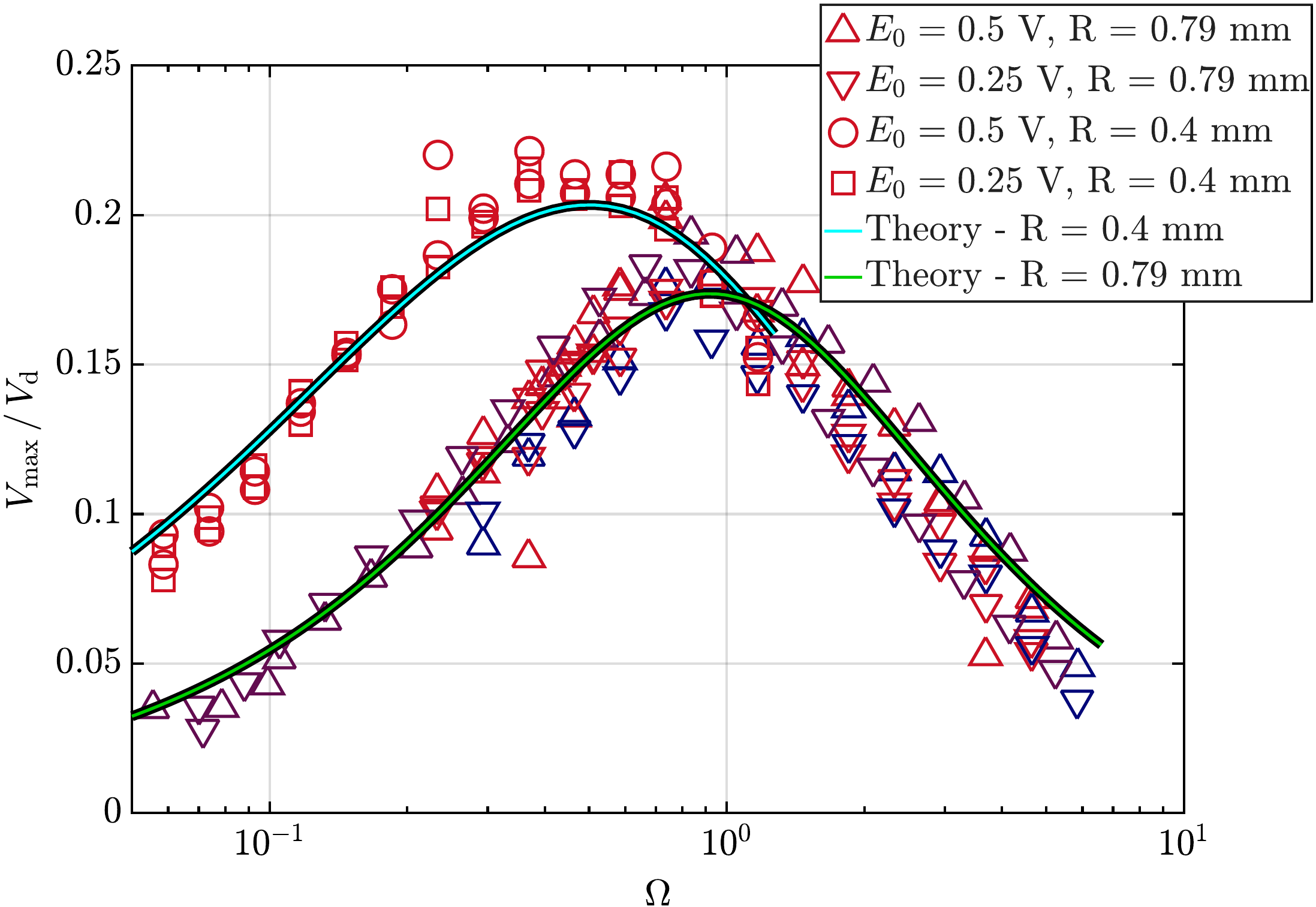}
    \caption{Dimensionless peak velocity, $V_\mathrm{max}/V_\mathrm{d}$ as a function of the dimensionless frequency $\Omega$. The solid line corresponds to the model predictions for a drop of typical length $L_\rmd = 9$ mm. Colorbar shared with Fig.~\ref{fig:expMeas}b.}
    \label{fig:dimensionless_results_comparison}
\end{figure}

\textit{Discussion} -- To compare the predictions of the model with the experiments, we assimilate the amplitude of the Fourier transform of the velocity, $|\hat{V}|$, to the peak amplitude of the velocity oscillations $V_\mathrm{max}$. In Fig.~\ref{fig:dimensionless_results_comparison} we show the experimental non-dimensional maximum velocity, $V_\mathrm{max}/V_\mathrm{d}$, corresponding to drops of different lengths, in the two capillary tubes, and excited at voltages $E_0 =$ \SI{0.25}{\volt} and \SI{0.5}{\volt}, as a function of a dimensionless frequency, $\Omega = \omega\mrho R^2/8\mmu$. The theoretical curves have been computed for a typical droplet length $L_\rmd = 9$ mm. Although these curves depend on the drop's length through $\mrho$ and $\mmu$, this dependency is weak for the drop and capillary lengths we use here.

The proposed model compares fairly well with the experimental measurements for over two decades ($\Omega = 0.05 - 6.0$) and for the two capillary radii. Notably, the model brings together data that is quite dispersed when plotted in dimensional variables (section S3). It is important to remark that only one fitting parameter has been used, the surface resistance of the drop, $R_\rms = $\SI{1.022}{\kilo\ohm} and $R_\rms = $\SI{1.253}{\kilo\ohm} for the setup with $R=0.79\,$mm and $R=0.4\,$mm respectively. The rest of the parameters have either been measured in the experiments or taken from the literature (Table \ref{tab:params} and section S6). The fitted values of the surface resistance are similar for the two geometries, indicating that the predictions are not sensitive to this parameter.

Interestingly, the proposed theory works well despite the assumption that both $C_\rms$ and $R_\rms$ are constant.
The fact that assuming constant surface capacity and drop resistance suffices to reproduce quantitatively the experimental results suggests that the model captures the physical ingredients of the electrocapillary-driven drop motion. This is consistent with the low voltage amplitudes that we use, always below the electrolysis threshold. Moreover, the lack of observed dependency of $R_\rms$ on the drop length suggests that most of this resistance occurs close to the menisci. Notice that both $C_\rms$ and $R_\rms$ depend on the applied voltage only for sufficiently large ones ($E_0 \gtrsim$ \SI{1}{\volt}) when irreversible surface reactions, such as electrolysis are important.

At this point we can explain why there is an optimal frequency that maximizes the peak velocity. Equation (\ref{eq:VelSolved}) relates linearly the dimensionless drop velocity to the electrical excitation in the Fourier space through two transfer functions: one mechanical and one electrical. On the one hand, the mechanical transfer function: $F_\mathrm{mech}(\omega) = \left(1 + \rmj \omega \mrho R^2 / 8 \mmu\right)^{-1}$, quantifies the effect of inertia in the drop motion.
This transfer function is well known in fluid mechanics: it appears in the flow of a viscous liquid inside a long tube subject to a pulsatile pressure gradient, the so-called Womersley flow, first studied in the context of blood flow in arteries \cite{womersleMethCalc55}. In fact, the dimensionless frequency $\Omega$ can be interpreted as the square of a Womersley number. 
At frequencies smaller than the inverse of the viscous time scale, $\omega \ll 8\mmu / \mrho R^2$, or $\Omega \ll 1$, viscous dissipation dominates over inertia, and the drop follows the electrocapillary pressure gradient with no delay and with an amplitude that is independent of the frequency. Conversely, for large frequencies, $\Omega \gg 1$, inertia comes into play, which makes hard for the liquids to follow the pulsatile electrocapillary pressure gradient. In this limit the transfer function behaves as $F_\mathrm{mech} \sim -\rmj\,\Omega^{-1}$. That is, the velocity follows the excitation with a $\pi/2$-delay and with an amplitude that decays inversely proportional to the frequency.

On the other hand, the electrical transfer function $F_\mathrm{elec}(\omega) = Z_\rmd/(Z_\rmd + Z_\rmc)$ is the ratio between the effective voltage that is actually modifying the drop's interfacial tension by electrocapillarity, $\Delta\phi_2-\Delta\phi_1$, and the excitation voltage applied to the whole circuit. To make the explanation more clear, we have also expressed $F_\mathrm{elec}$ in terms of the drop impedance $Z_\rmd = (1/R_s + \rmj \pi \omega R^2 C_\rms)^{-1}$, and the capacity $C_\rmd$ in terms of the surface capacity $C_\rms$. The circuit impedance without drop, $Z_\rmc$, follows the typical response of an electrochemical cell, diverging to infinity as a negative power law for $\omega \rightarrow 0$ and decaying to a constant value for $\omega \rightarrow \infty$ \cite{lazanasElecImpe23} (Section 1 \cite{suppmat}). Consequently, at low frequencies most of the voltage is lost on this impedance, and little is left to modify the drop's interfacial tension.

Besides modulating the amplitude of the velocity oscillations, the capacitive nature of the electrical transfer function also introduces a phase-lag between the excitation voltage and the voltage drop across the menisci, which eventually drives the motion. The phase-lag observed between the excitation voltage $E(t)$ and the velocity $V(t)$ in Fig.~\ref{fig:expMeas}a is a combination of the phase-lag induced by each transfer function.

The fact that the mechanical and electrical transfer functions exhibit opposite behaviors explains the existence of an optimal frequency at the cross-over between them. We remark here that our model not only explains the existence of the optimal velocity, but also predicts its value and at which frequency it is reached.
Notice that mechanism behind the existence of a maximum is different from a resonance since, if not externally forced, the drop does not oscillate on its own.

Examination of the expressions for $F_\mathrm{mech}$ and $F_\mathrm{elec}$ explains why the dimensionless theoretical curves in Fig.~\ref{fig:dimensionless_results_comparison}, corresponding to different radii, do not collapse on each other, and neither do the experiments. In both transfer functions the frequency appears explicitly multiplying the square radius, $\omega R^2$. Thus, the dimensionless frequency $\Omega = \bar{\rho} \omega R^2 / 8 \bar{\mu}$ could in principle collapse the data for both radii. However, $\omega$ also appears implicitly through the circuit impedance, $Z_\rmc(\omega)$, which depends very little on $R$ (Fig.~\ref{fig:impedances}). Consequently, the different dependencies on $R$ of the transfer functions and $Z_\rmc$ make impossible to derive any dimensionless frequency capable of collapsing curves for different capillary radii.

We can also use the theory to infer the role, or lack of it, of other physical effects that are commonly found in similar microfluidic systems. First, the main source of dissipation in the flow are viscous stresses in the bulk. This is a consequence of the existence of a thin electrolyte film between the drop and the tube, which avoids the formation of a triple contact line solid-electrolyte-drop. When a triple contact line exists, the dissipation in its vicinity can become dominant \cite{dolletTranExpo20, ruiz-gutLongCros22}. In fact, in the experiments at the lowest frequencies ($f \approx $ \SI{0.1}{\hertz}) we start to observe nonlinear effects, when the drop speed approaches zero at each cycle, that can be attributed to localized wetting of the walls by the metal. We discuss this marginal effect in \cite{suppmat} (Section S3).

A second effect that we must discuss is the role of shear stresses. The electrocapillary-driven surface tension gradient creates shear stresses analog to Marangoni ones at the drop interface \cite{schnitzer2013electrokinetic}. However, the proximity of the interface to the tube wall, where no slip is imposed, makes the interfacial velocity very small compared to that in the bulk. If the drop were not confined inside a capillary, shear stresses would contribute to the drop motion and deformation. Such a situation has been observed, for instance, when an EGaIn slug flows into a large electrolyte reservoir from a capillary and atomizes into drops in the presence of an electric field \cite{tangSteeLiqu15}.

In summary, we study here the response of an EGaIn drop in a capillary tube filled with an electrolyte subject to an oscillatory electrical excitation. This system is a model that allows us to elucidate the physical mechanisms that determine the response time of a liquid metal drop to a time-varying electrical excitation. 
Although we apply our theory to this particular system, it stands to reason that its predictions can be generalized to any drop of liquid metal immersed in an electrolyte as long as: (a) the drop is longer than the radius of the capillary, (b) the combination voltage/electrolyte is such that irreversible surface reactions do not take place, and (c) the drop does not stick to the walls.
The problem we address here is not only of fundamental interest, but also relevant to modern practical applications of Gallium-based alloys. Despite the notable advance in Gallium-based electronic circuits \cite{babatain2025programmable}, to the best of our knowledge, an analysis of transients based on first principles has not been done. It is worth pointing out that in current microfluidic electrical circuits the response times are usually long, of the order of the second. Thus, there is plenty of room for optimization. We believe that the model we propose in this Letter will be useful for this purpose.

\begin{acknowledgments}
The authors are grateful to Detlef Lohse, Alvaro Marin and Francisco Higuera for their support and enthusiasm on this project, shown through very fruitful and estimulating discussions. J.R.R., J.O.M., and A.G.A. acknowledge funding by MICIU/AEI/10.13039/501100011033. J.R.R acknowledges financial support from Grant No. PID2023-146809OB-I00 funded by ERDF/UE and Grant No. PID2020-114945RB-C21. J.O.M and A.G.A. acknowledge partial funding by projects HE 101192080 (6G-LEADER) and by PID2023-147305OB-C31 (SOFIA-AIR): funded by the European Union. Views and opinions expressed are however those of the author(s) only and do not necessarily reflect those of the European Union. Neither the European Union nor the granting authority can be held responsible for them. This work is supported by the grant from Dutch Research Council (NWO), VI.C.222.037.
\end{acknowledgments}

%



\newpage

\setcounter{figure}{0}
\renewcommand\thefigure{S\arabic{figure}}
\renewcommand\thetable{S\arabic{table}}
\renewcommand\thesubsection{S\arabic{subsection}}
\setcounter{equation}{0}
\renewcommand\theequation{S\arabic{equation}}

\onecolumngrid

\noindent \rule{\columnwidth}{0.5pt}

\setcounter{page}{1}

\section{Supplemental Material:\\Maximum Velocity of a Liquid Metal Drop Driven by a\\Time-Dependent Electrocapillary Excitation}

\setcounter{subsection}{1}
\subsection{Section \thesubsection: Determination of the circuit impedance with and without drop}

The mathematical model developed in the main text needs the impedance of the circuit without drop, $Z_\mathrm{c}$, to compute the electrical transfer function $F_\mathrm{elec}$. This impedance has been determined experimentally in a dedicated campaign of experiments without drop. To that end, the circuit has been filled with the electrolyte solution and the same excitation voltages used to move the drop have been applied at the electrodes. The impedance has been determined at each combination voltage-frequency by minimizing the following quadratic error:
\begin{equation}
    \varepsilon^2 = \sum_{n}\left[\left(\frac{E_n}{E_0}\right)^2 + \left(-\frac{I_n \left|Z_\mathrm{c}\right|}{E_0\sin\varphi} + \frac{E_n\cos\varphi}{E_0\sin\varphi}\right)^2 - 1\right]^2.
\end{equation}

\begin{figure}[b]
    \centering
    \includegraphics[width=0.7\linewidth]{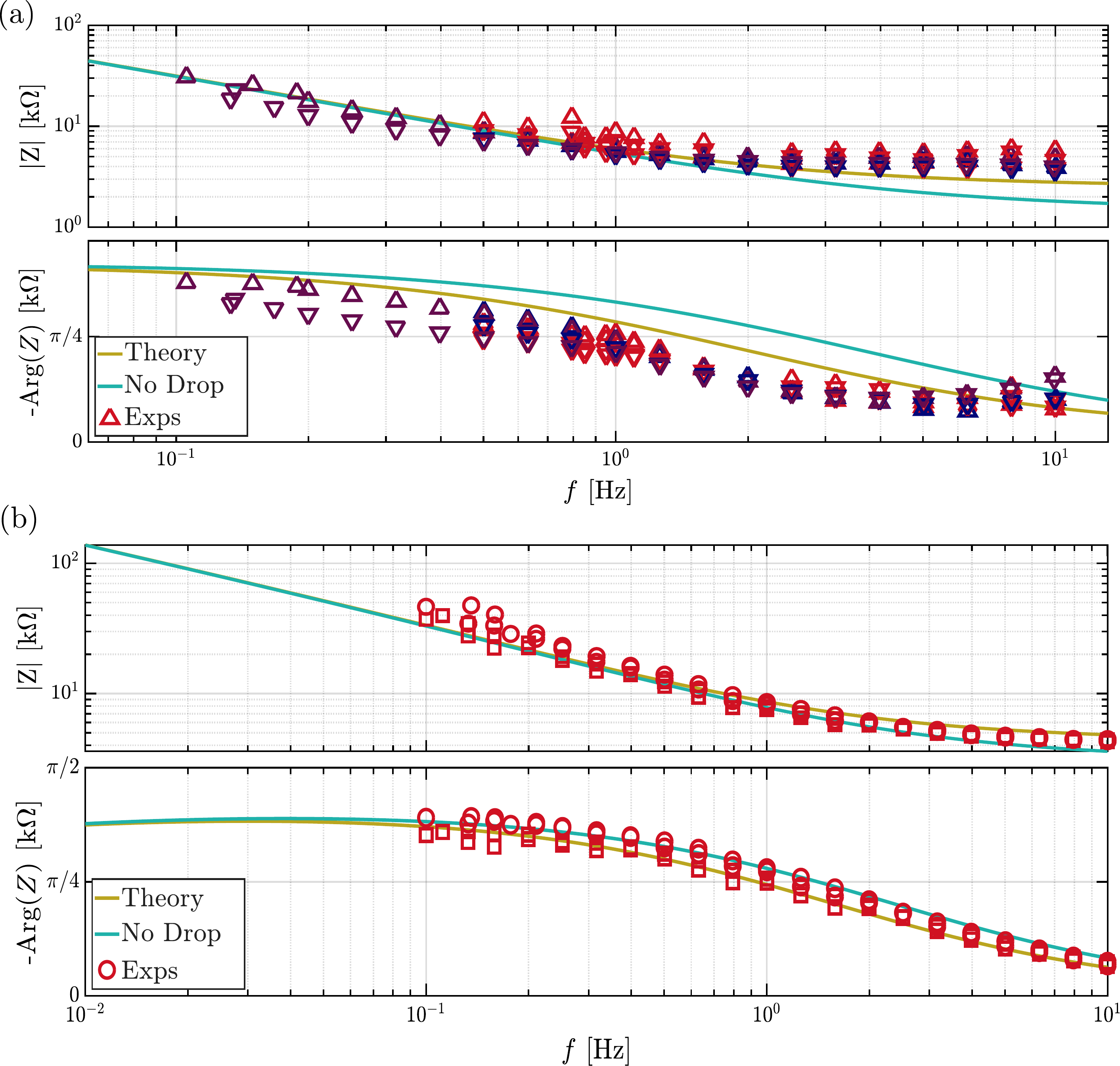}
    \caption{Comparison between the impedance predicted by the model (brown line) and the experimental one with drop (symbols). The cyan line is a fitting to the experimental impedance of the circuit with no drop, $Z_\mathrm{c}$. Here, (a) stands for the experimental setup with $R = 0.79\,$mm, and (b) for $R = 0.4\,$mm.}
    \label{fig:impedances}
\end{figure}

Here, $E_n$ and $I_n$ are simultaneous measurements of the excitation voltage and the current taken during each experiment. These samples were restricted to the time interval excluding the first and the last cycle of the current oscillations, to avoid transients. The same procedure has then been applied to the current measured during the experiments with drop. Figure \ref{fig:impedances} shows the modulus, $\left|Z\right|$ and the phase $\mathrm{Arg}(Z)$ of the measured impedances with and without drop for different frequencies and for $E_0 = $ \SI{0.25}{\volt} and $E_0 = $ \SI{0.5}{\volt}. Markers represent experiments with drop whereas the solid blue line is a polynomial fit in the variable $\log_{10}(f)$ done for the impedances without drop. We also show the predictions of our theoretical model (brown line), given by:
\begin{equation}
    Z_\mathrm{d}(\omega) = Z_\mathrm{c}(\omega) + \left(\frac{1}{R_s} + \rmj\omega\pi R^2 C_\mathrm{s}\right)^{-1}.
\end{equation}
The agreement between the experimental and the theoretical impedance with drop is reasonably good, given that there are no free parameters fitted for this comparison, as $C_\mathrm{d}$ is taken from experiments reported in the literature \cite{khanGianSwit14} and $R_\mathrm{s}$ has already been obtained by fitting the theoretical expression for the velocity to match the experiments (see Table \ref{tab:params} for the values of these parameters).

To conclude this section, we point out that an additional effect reducing the electrical transfer function is expected to occur at frequencies of the order of or larger than $f_\mathrm{c-o} \approx 1/\pi R_\rms C_\rmd \approx $ \SI{224}{\hertz}. At these frequencies, the impedance of the surface capacitors decays to zero as $Z_\rmd \sim f^{-1}$, and the drop loses its ability to sustain an electrocapillary voltage. This regime where both transfer functions decay as $\omega^{-1}$ is hard to observe experimentally, as the drop velocities would be simply too small. However, for the same reason, is not a regime of practical interest.

\begin{table}[b]
    \centering
    \begin{tabular}{|c l|c|c|c|}
        
        \multicolumn{2}{c}{\textbf{Parameter}} & \multicolumn{1}{c}{\textbf{Value}} & \multicolumn{1}{c}{\textbf{Units}} & \multicolumn{1}{c}{\textbf{Ref.}} \\ \hline 
        \multicolumn{5}{l}{\textbf{Electrolyte (1 M NaOH water solution)}}\\\hline\hline
        $\epsilon_\rme$ & Permittivity  & 64.42 @25$^\circ$C & - & \cite{buchnerDielRela99} \\ 
        $\sigma_\rme$ &Conductivity & 17.88 @25$^\circ$C & $\Omega^{-1} \rm{m}^{-1}$ & \cite{buchnerDielRela99} \\ 
        $\rho_\rme$ &Density & 1040 & kg$^{-3} \rm{m}^{-1}$ & \cite{buchnerDielRela99} \\
        $\mu_\rme$ &Viscosity & 1.4 & mPa s & \cite{buchnerDielRela99} \\
        \hline 
        \multicolumn{5}{l}{\textbf{EGaIn}}\\\hline\hline
        $\rho_\rmd$ &Density & 6250 & kg$^{-3} \rm{m}^{-1}$ & \cite{xuEffeOxid12} \\
        $\mu_\rmd$ &Viscosity & 1.99 & mPa s & \cite{dickeyEuteGall08} \\
        $\gamma_0$ &Surf. tension @pzc & 0.51 & N m$^{-1}$ & \cite{khanInflWate14} \\
        $C_\mathrm{s}$ &Surf. capacity @pzc & 30 & $\mu$F cm$^{-2}$ & \cite{khanInflWate14}(*) \\
        $C_\mathrm{d} = 2\pi R^2 C_\mathrm{s}$ & Meniscus capacity & 1.176 & \SI{}{\micro\farad} & - \\
        $\phi_0$ &Voltage @pzc & -0.91 & V & \cite{khanInflWate14} \\
        \hline 
        \multicolumn{5}{l}{\textbf{Capillary dimensions}}\\\hline\hline
        $L_\mathrm{t}$ & Capillary length & 110 & mm & - \\
        $R$ & Capillary inner radius & 0.79 & mm & - \\
         &  & 0.4 & mm & -\\ 
        \hline  \multicolumn{5}{l}{\textbf{Fitted parameters}}\\\hline\hline
        $R_\mathrm{s}$ & Surface resistance (with R = 0.79)& 1022 & $\Omega$ & - \\
          &  ~~~~~~~~~~~~~~~~~~~~~~~~~(with R = 0.4) & 1253 & $\Omega$ & - \\
        \hline 
    \end{tabular}
    \caption{Experimental parameters used to obtain the results presented in this manuscript. The parameters for which no reference is given have been obtained as part of this work. (*) The surface capacity has been taken as that without surface oxidation observed in Fig. 1E of Ref. \cite{khanInflWate14}.}
    \label{tab:params}
\end{table}

\setcounter{subsection}{2}
\subsection{Section \thesubsection: Experimental Workflow \& Video processing}

\begin{figure}[t]
    \centering
    \includegraphics[width=0.55\linewidth]{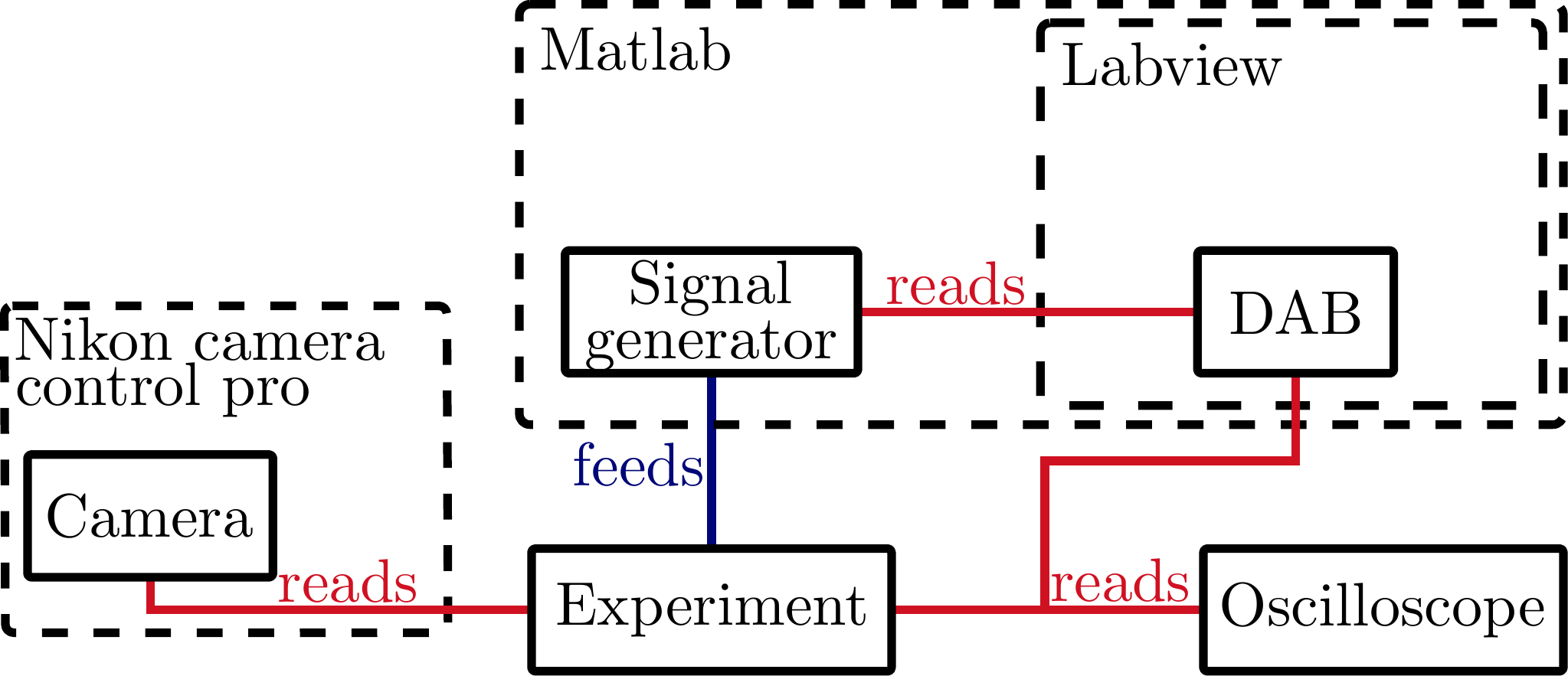}
    \caption{Experimental setup and workflow control.}
    \label{fig:workflow}
\end{figure}

The camera is controlled using Nikon Camera Control Pro 2, saving video readings into a dedicated text file. The signal generator is remotely controlled by an in-house script written in Matlab. The script generates the excitation waveform that the signal generator will produce. Also, the script randomizes the order of the excitation frequency in the different experiments to avoid memory effects. The digital acquisition board (DAB) (National Instruments NI-6008) is controlled using Labview. It requires a trigger, generated by the same script mentioned above, to record both the input signal and the voltage drop at the measurement resistor $R_\mathrm{m}$. This trigger ensures the synchronization between the acquired signals, $E(t)$ and $I(t)$. The current flowing through the device can be readily computed as  $I = V_{\rm m} / R_{\rm m}$. Finally, the synchronization between the videos and the electrical measurements of $E$ and $I$ is done manually for each experiment. The overall measurement process is summarized in Fig~\ref{fig:workflow}.

The post-processing is done using a home-made code written in Matlab. Firstly, data files are loaded into a single variable. For each video frame, we carried out the following process:

\begin{enumerate}[noitemsep]
    \item Crop frame and bit depth adjustment (if necessary)
    \item \textit{Edge detection}: after the image is binarized, the drop can be easily differentiated from the background thanks to the high contrast, using a gray-level threshold. Using the position of the drop position in the binarized image as initial guess, we then interpolate to sub-pixel resolution the position of the drop edge in the original gray-level image. Drop length is computed subtracting edge positions, and the drop position is calculated as the average between the edges.
    \item \textit{Velocity computation}: To obtain the velocity, differentiation is performed using finite differences with a time step equal to twice the frame rate. Moreover, drop central position curves were smoothened before differentiating using a Gaussian kernel and variable window size, depending on the frequency of the corresponding experiment.
    \item \textit{Current computation}: current is computed dividing the voltage measured at the calibrated resistor by its resistance, which was 220 $\Omega$ in all our experiments. This value is smoothened using a moving average filter of 20 points and compensated subtracting its mean.
\end{enumerate}

\textit{Frequency resolution:} The highest excitation frequency we use in our experiments, $f = 10$ Hz, is much smaller than the frame rate at which the movies are acquired, 120 Hz (fps). Following Nyquist theorem, we would be able to resolve frequencies as high as 60 Hz in our experiments, which are still very large compared to our maximum excitation frequency. The large sampling rate compared to the excitation frequency allows us to implement low-pass filters to compute the drop speed in the experiments, as described in Section S2.\\
 
\textit{Velocity resolution:} The spatial resolution in our experiments is of the order S = 60 pixels/mm. At the typical peak speeds of the drops ($V_\mathrm{max} \sim 20$ mm/s), the drop’s meniscus translates a distance $d = V_\mathrm{max}/\mathrm{fps} \sim 0.17$ mm, that is, about 10 pixels. This allows us to accurately determine the displacement between consecutive frames. In fact, we use a sub-pixel interpolation technique that allows us to determine the position of the meniscus with a resolution better than 1 pixel. That is, at maximum speeds, the typical displacement between two consecutive frames can be measured with a resolution of at least a few units per cent.\\

\textit{Averaging of the peak velocity:} the number of oscillation cycles, $N_c$, used to get the average peak velocity depends on the excitation frequency. For frequencies smaller than 1 Hz: $N_c = 10$. For 1 Hz $\leq f <$ 2.5 Hz, $N_c = 15$. For 2.5 Hz $\leq f <$ 6 Hz, $N_c = 20$. For 6 Hz $\leq f <$ 7 Hz, $Nc = 25$. For 7 Hz $\leq f \leq$ 10 Hz, $N_c = 30$. Notice that we do not include the data from the first and the last oscillation cycle in the average, to exclude the effect of transients. Also, we get two values for the average per cycle: the maximum and the minimum peaks.\\

\setcounter{subsection}{3}
\subsection{Section \thesubsection: Non-harmonic velocity time evolution at low excitation frequencies}

\begin{figure}[h]
    \centering
    \includegraphics[width=0.65\linewidth]{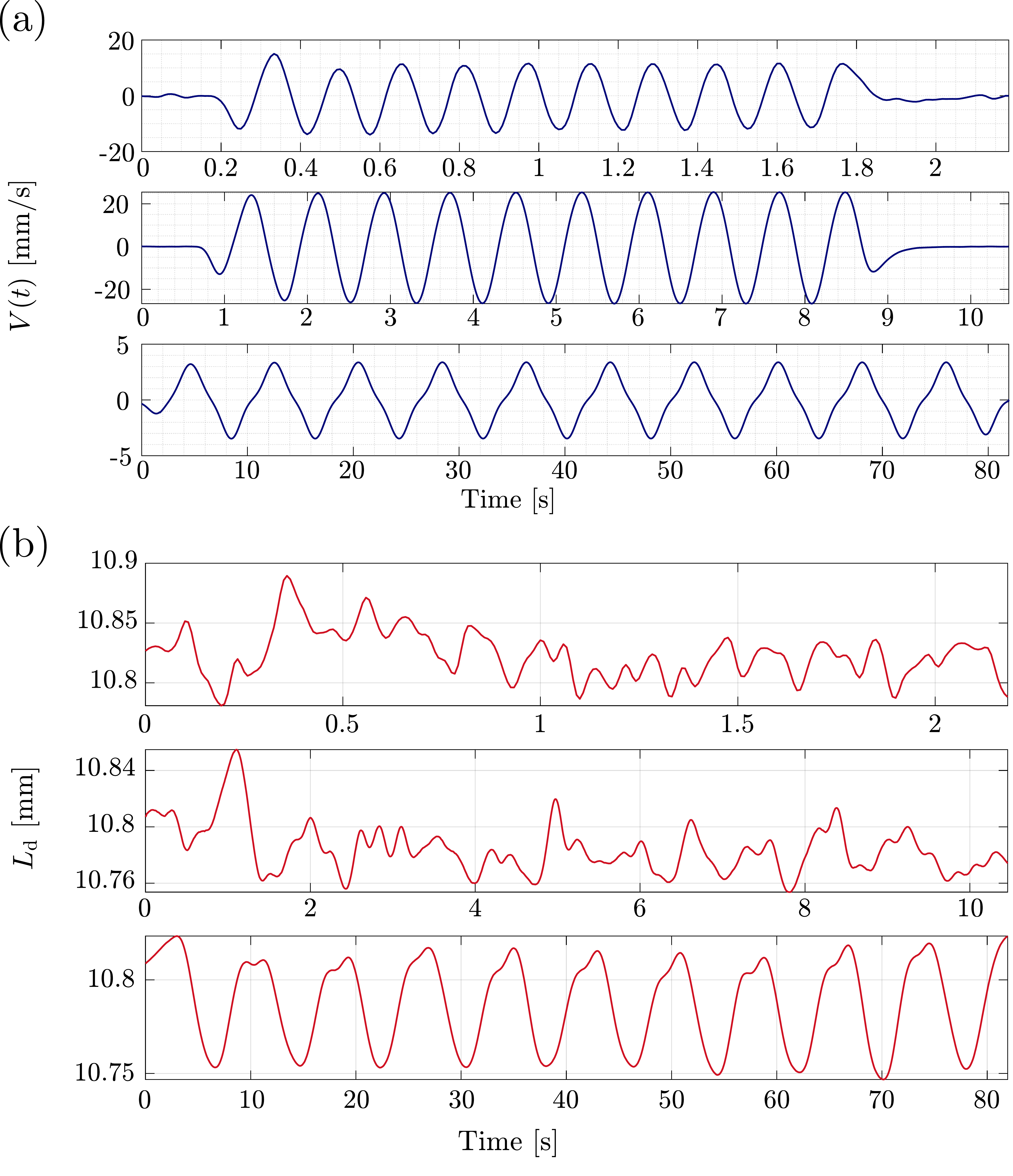} 
    \caption{ \label{fig:drop_velocity_several_frequencies}(a) Drop velocity of three experiments excited at $f = $ 6.31 Hz, 1 Hz and 0.125 Hz for an excitation amplitude $E_0 = 0.5$ V. (b) Time evolution of the drop length for the same experiments shown in (a).} 
\end{figure}

Although both the electrical excitation that we use to move the drop $E(t)$ and the circuit current $I(t)$ follow pure harmonic oscillations, the drop velocity exhibits higher harmonics at the lowest frequencies that we explore in our experiments ($f \approx 0.1$ \SI{}{\hertz}). To illustrate the nature of these oscillations, we show in Figure \ref{fig:drop_velocity_several_frequencies} the time evolution of the drop velocity for a drop of length $L_\rmd = $ \SI{10.8}{mm} excited with $E_0 = $\SI{0.5}{\volt} at frequencies $f = $ \SI{6.3}{\hertz} (top), \SI{1}{\hertz} (center), and \SI{0.125}{\hertz} (bottom). Removing the first and last oscillation cycles, where transient effects are noticeable, the top and center plots reveal that the drop follows purely sinusoidal oscillations. However, at the bottom plot, we can see how the velocity changes more slowly when its value is close to zero. We hypothesize that this effect is due to the sticking of the metal to the capillary's surface.

Indeed, at frequencies such that $\Omega \gtrsim 1$, inertia is important. Thus, during the part of the oscillation cycle when the electrocapillary pressure gradient is small the drop keeps translating by inertia, which slows down the drainage of the liquid film that separates its surface from the wall and thus inhibits wetting. Conversely, during the part of the cycle when the velocity is small, the electrocapillary pressure gradient is large, which makes the drop overcome any friction force arising from its contact with the walls. This situation is very different at low frequencies, where inertia is irrelevant. During the part of the oscillation cycle when the electrocapillary pressure gradient is small the drop velocity can be small enough for the film separating the drop and the wall to drain and therefore to let the drop stick to the wall. Then, it takes a given electrocapillary pressure gradient to make this union yield and let the drop translate again (see section S3 for more details). We remark that, even in these cases, the contact angle between the drop menisci and the wall is as close to 180$^\circ$ as we can measure. Thus, the flow is still driven by electrocapillary pressure gradients, not by changes in the contact angle as is the case in electrowetting on a dielectric \cite{lu2007diffuse}, which requires much larger voltages.

Notice that this stick-yield behavior has no effect on the peak velocity that the drop reaches in each cycle, which is what we use to compare our theory with. Indeed, if inertia is unimportant, the drop reaches the velocity at which the electrocapillary forces balance viscous stresses instantaneously. This explains why our theory does not need to account for these second-order effects to successfully reproduce the curves $V_\mathrm{max}$ vs. $f$.

\setcounter{subsection}{4}
\subsection{Section \thesubsection: Additional plots of the maximum velocity and current}
\label{sec:section_plots_allR}

\begin{figure}[t]
    \centering
    \includegraphics[width=\textwidth]{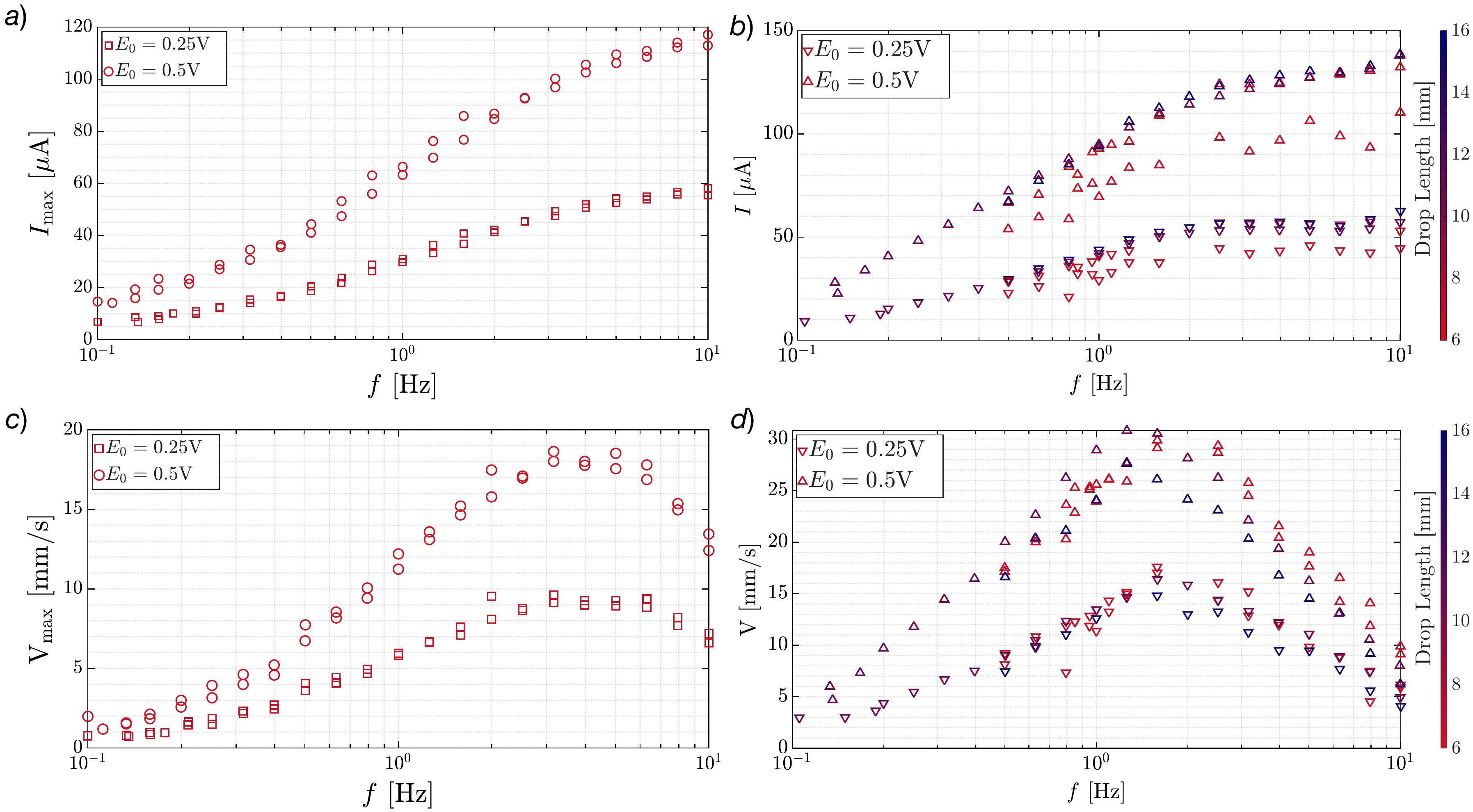}
    \caption{\label{fig:measThinCap}Peak current $I_{\rm max}$ for the capillary with $R =$ \SI{0.4}{\milli\meter} (a) and $R =$ \SI{0.79}{\milli\meter} (b). Peak velocity $V_\mathrm{max}$ for the capillary with $R =$ \SI{0.4}{\milli\meter} (c) and $R =$ \SI{0.79}{\milli\meter} (d). Results in (a) and (c) correspond to two different drops of length $L_\rmd =$ \SI{6}{\milli\meter}. The size of the drops in panels (b) and (d) corresponds to the color code of the color bars on the right.}
\end{figure}

We show here in Fig~\ref{fig:measThinCap} the average peak values of the current and velocity for drops in the two capillary tubes. The top panels (a and b) display the peak current, whereas the bottom ones (c and d) represent the peak velocity. Figures representing the same magnitude are placer side-by-side so the different vertical scales can be better compared. The code used to generate the plots and reproduce the analysis, along with the complete underlying dataset, is openly available in Zenodo at~\cite{dataZenodo}.

\begin{figure}[t]
    \centering
    \includegraphics[width=0.65\linewidth]{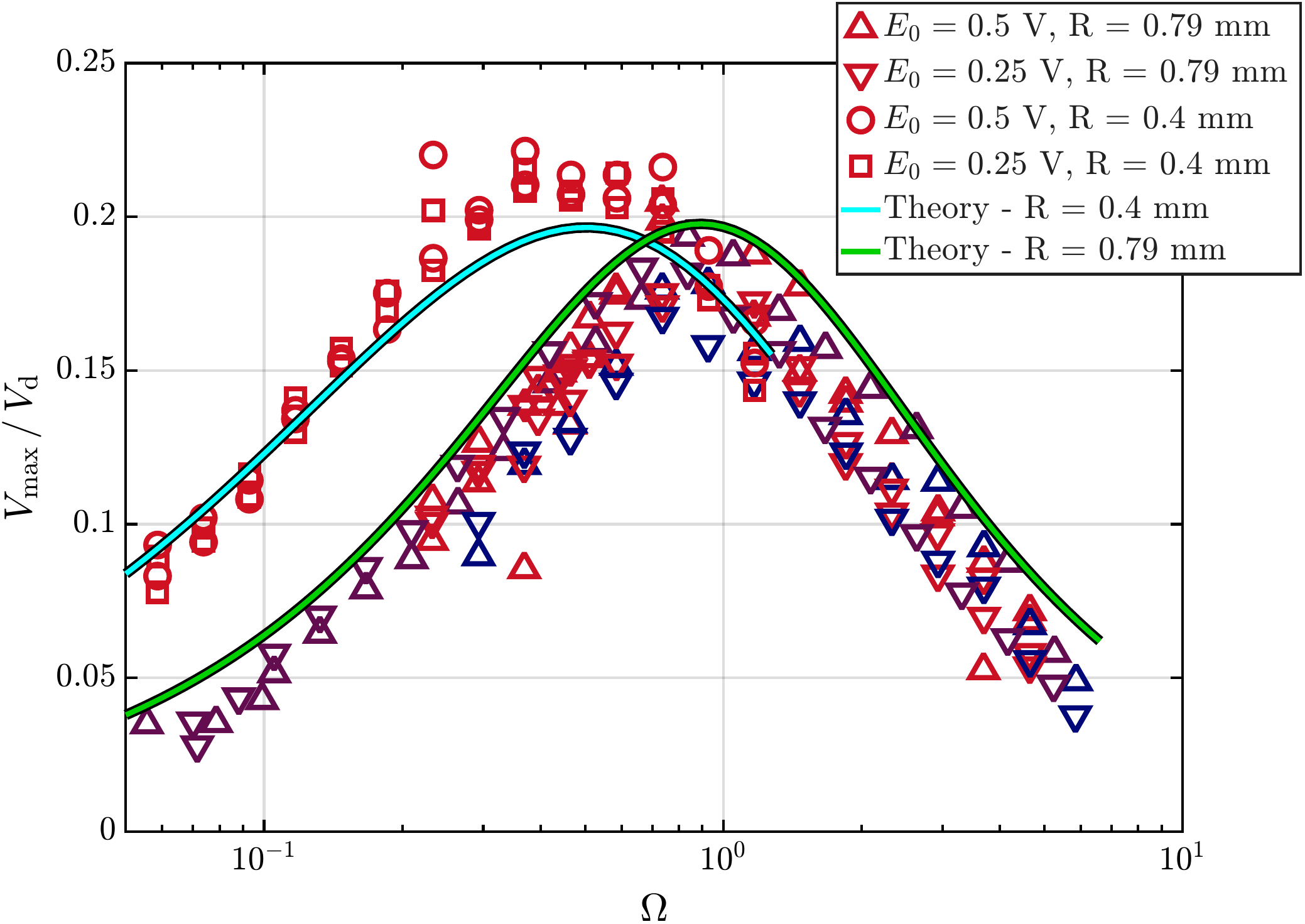}
    \caption{\label{fig:VmaxVd_vs_Omega_same_Rs}Dimensionless peak velocity, $V_\mathrm{max}/V_\mathrm{d}$ as a function of the dimensionless frequency $\Omega$. The solid line corresponds to the model predictions using the same surface resistance $R_\rms =$ \SI{1200}{\ohm}.}
\end{figure}
\setcounter{subsection}{5}
\subsection{Section \thesubsection: Additional details on the derivation of the mathematical model}
\subsubsection{Derivation of the differential equation for the flow velocity, $V$}

\begin{figure}[H]
    \centering
    \includegraphics[width=0.6\linewidth]{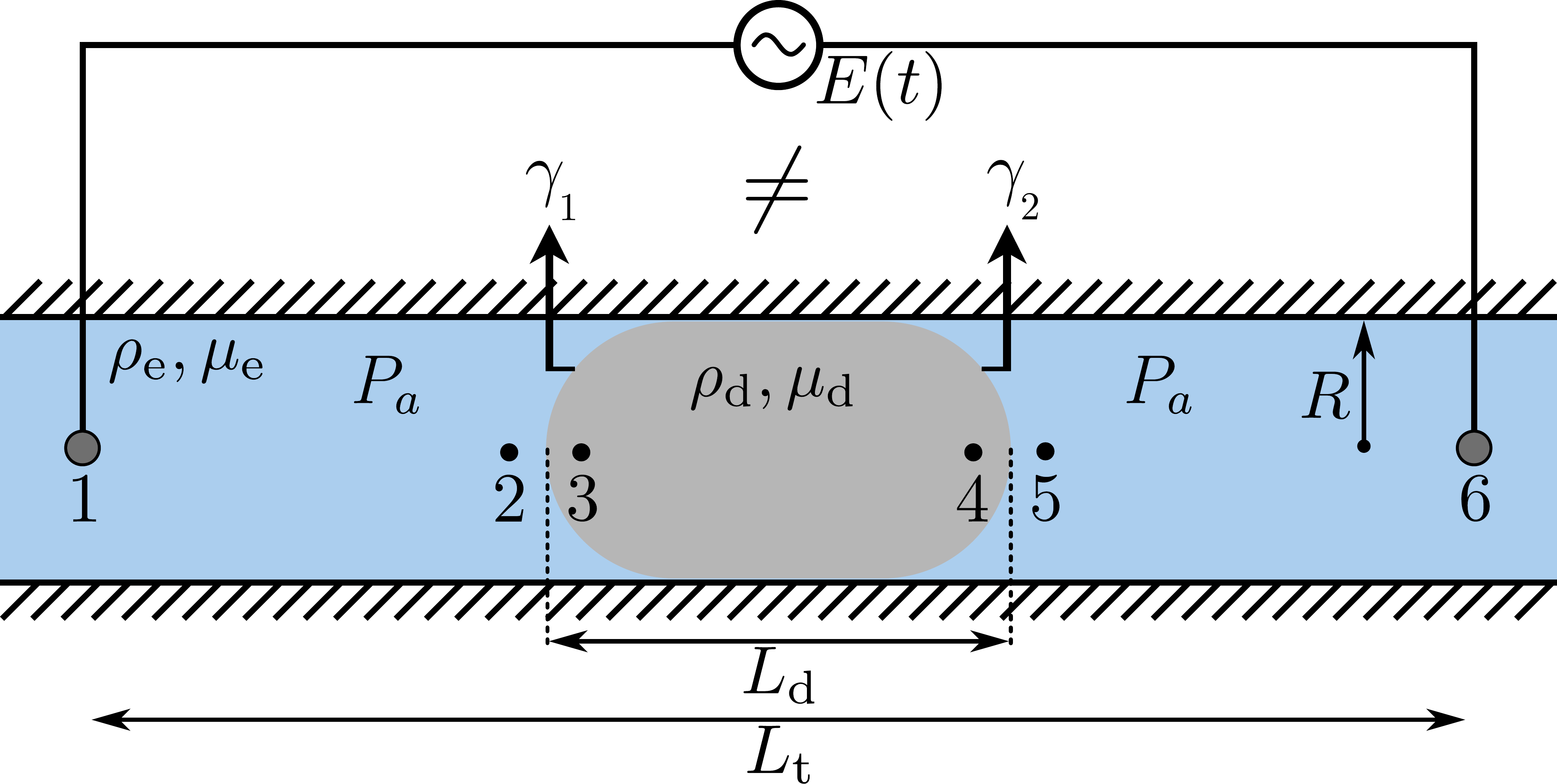}
    \caption{Sketch of the flow with the definition of the control points to formulate the mathematical model.}
    \label{fig:sketch}
\end{figure}

We start by writing the equation for the mean velocity in the one-directional flow of a viscous liquid in a straight pipe \cite{quereInerCapi97} for the three legs of the flow: left electrolyte (1-2), drop (3-4), and right electrolyte (5-6), Fig. \ref{fig:sketch}:
\begin{eqnarray}
    \rho_\rme L_{1-2} \dfrac{\rmd V}{\rmd t} & = & P_a-P_2 - \, \dfrac{8\mu_\rme V L_{1-2}}{R^2}\\
    \rho_\rmd L_\rmd  \dfrac{\rmd V}{\rmd t} & = & P_3-P_4 - \, \dfrac{8\mu_\rmd V L_\rmd}{R^2}\\
    \rho_\rme L_{5-6} \dfrac{\rmd V}{\rmd t} & = & P_5-P_a - \, \dfrac{8\mu_\rme V L_{5-6}}{R^2}.
    \label{eq:quere_three_legs}
\end{eqnarray}
Here, we have already made use of the fact that the length between points 3 and 4 is equal to the length of the drop, $L_\rmd$, and that the pressure at points 1 and 6 is nearly the ambient one, $P_a$. Notice that the velocity $V$ must be the same in the three legs by mass conservation.

Adding the three equations together,
\begin{equation}
    \left(\rho_\rme (L_\rmt-L_\rmd) + \rho_\rmd  L_\rmd\right) \dfrac{\rmd V}{\rmd t} = (P_3-P_2) - (P_4-P_5) + \, \dfrac{8 V}{R^2}\left(\mu_\rme (L_\rmt-L_\rmd) + \mu_\rmd L_\rmd\right).
\end{equation}
At this point, we define $\bar{\rho} = \left(\rho_\rme (L_\rmt-L_\rmd) + \rho_d L_\rmd\right)/L_\rmt$ and $\bar{\mu} = \left(\mu_\rme (L_\rmt-L_\rmd) + \mu_\rmd L_\rmd\right)/L_\rmt$. Also, denoting by $\gamma_1$ and $\gamma_2$ the interfacial tensions between the drop and the electrolyte at menisci 2-3 and 4-5 respectively, 
\begin{equation}  
     P_3 - P_2 = \dfrac{2\gamma_{1}}{R}~~,~~P_4 - P_5 = \dfrac{2\gamma_{2}}{R}.
    \label{eq:YoungLaplace}
\end{equation}
Strictly speaking, the asymmetric interfacial tensions caused by the electrical excitation lead to an asymmetry in the shape of the menisci. However, for this asymmetry to be observed we need the change in the surface tension $\Delta\gamma$ to be of the order of the surface tension at zero voltage, $\gamma_0$. Since $\gamma_0$ is large for liquid metals ($\gamma_0 \approx$ 500 mN/m), the changes in the curvature of the menisci are not visible for voltages smaller than the threshold of occurrence of surface reactions.\\

Combining the above expressions we finally arrive at:
\begin{equation}
    \bar{\rho}L_\rmt\frac{\mathrm{d}V}{\mathrm{d}t} = \dfrac{2 \, (\gamma_1-\gamma_2)}{R} - V\dfrac{8\mmu L_\rmt}{R^2}.
    \label{eq:surfs}
\end{equation}
Equation (\ref{eq:velocity_interfa}) is ultimately obtained by dividing this equation by $\mrho L_\rmt$.

The next step is to express the interfacial tension difference $\gamma_1 - \gamma_2$ as a function of the electrocapillary voltage difference $\Delta\phi_1 - \Delta\phi_2$. Using Lipmann's law:
\begin{equation}
    \gamma_1 - \gamma_2 = -\frac{1}{2}C_\mathrm{s}\left(\Delta\phi_1 - \phi_0\right)^2 + \frac{1}{2}C_\mathrm{s}\left(\Delta\phi_2 - \phi_0\right)^2 = C_\mathrm{s}\phi_0\left(\Delta\phi_2 - \Delta\phi_1\right) + \frac{1}{2}C_\mathrm{s}\left((\Delta\phi_2)^2-(\Delta\phi_1)^2\right).
\end{equation}
This expression can be further simplified by using the fact that $\Delta\phi_2 + \Delta\phi_1 = 0$ at all times. Indeed, due to the symmetry between the two menisci, the voltage jumps obey the following differential equations:
\begin{equation}
    I_\rmd = C_\rmd\frac{\rmd\Delta\phi_2}{\rmd t} = -C_\rmd\frac{\rmd\Delta\phi_1}{\rmd t},
\end{equation}
where $I_\rmd$ is the current that flows through the drop. Consequently,
\begin{equation}
    \gamma_1 - \gamma_2 = C_\mathrm{s}\phi_0\left(\Delta\phi_2 - \Delta\phi_1\right),
\end{equation}
and finally
\begin{equation}
    \frac{\mathrm{d}V}{\mathrm{d}t} = \dfrac{2C_\mathrm{s}\phi_0\left(\Delta\phi_2 - \Delta\phi_1\right) }{\bar{\rho} L_\rmt R} - V\dfrac{8\mmu}{\mrho R^2}.
    \label{eq:dVdt_vs_Dphi12}
\end{equation}

\subsubsection{Expression for the velocity in the Fourier space}

The electrocapillary voltage difference $(\Delta\phi_1-\Delta\phi_2)$ is more easily computed in the Fourier space. The reason is because the impedance of the circuit without drop, $Z_\rmc$, is obtained experimentally in the Fourier domain \cite{lazanasElecImpe23} and it does not admit a simple expression in the time domain. Consequently, determining the current $I(t)$, needed to compute the electrocapillary voltage difference, involves performing a complicated convolution integral. This would complicate the theoretical model and, more importantly, make it harder to interpret it physically, which is one of our main objectives.\\

For the above reasons we perform our calculations from now on in the Fourier space, starting by transforming Equation (\ref{eq:dVdt_vs_Dphi12}):
\begin{equation}
    \mathrm{j}\omega \hat{V}=\frac{2 C_\rms \phi_0}{\mrho L_\rmt R}\left(\Delta\hat{\phi}_2-\Delta\hat{\phi}_1\right) - \frac{8\mmu}{\mrho R^2}\hat{V},
\end{equation}
or, isolating the velocity,
\begin{equation}
    \hat{V} = \frac{C_\rms \phi_0 R}{4 \mmu L_\rmt}\left(1 + \mathrm{j}\frac{\omega \mrho R^2}{8\mmu}\right)^{-1}\left(\Delta\hat{\phi}_2-\Delta\hat{\phi}_1\right).
    \label{eq:velocity_fourier_SM}
\end{equation}\\

In view of the proposed drop equivalent circuit (Figure \ref{fig:elec_cir}b), the electrocapillary voltage can be obtained in the Fourier space as follows:
\begin{equation}
    \Delta\hat{\phi}_2-\Delta\hat{\phi}_1 = Z_\rmd \hat{I},
    \label{eq:vdifs}
\end{equation}
where $Z_\rmd$ is the drop impedance,
\begin{equation}
    Z_\rmd = R_\rms \myparal (C_\rmd/2) = \dfrac{1}{1/R_\rms + \rmj\omega C_\rmd/2}.
    \label{eq:zdrop}
\end{equation}
The current is finally related to the excitation voltage,
\begin{equation}
    \hat{I} = \frac{\hat{E}}{Z_\rmc + Z_\rmd},
\end{equation}
thus
\begin{equation}
     \Delta\hat{\phi}_2 - \Delta\hat{\phi}_1 = \frac{Z_\rmd}{Z_\rmc + Z_\rmd}\hat{E} = \left(1 + Z_\rmc \left( \dfrac{1}{R_\rms} + \dfrac{j\omega C_\rmd} {2}\right)\right)^{-1} \hat{E}.
\end{equation}
Introducing this expression into (\ref{eq:velocity_fourier_SM}), and nondimensionalizing the excitation voltage with its amplitude $E_0$, we finally arrive at
\begin{equation}
    \Hat{V} = \frac{C_\rms \phi_0 R E_0}{4\mmu L_\rmt} \left(1 + \rmj\frac{\omega \mrho R^2}{8 \mmu}\right)^{-1} \left(1 + Z_{\rm c}\left(\dfrac{1}{R_\rms} + \rmj\omega\pi R^2 C_{\rm s}\right)\right)^{-1} \dfrac{\Hat{E}}{E_0},
\end{equation}
which is Eq.~(\ref{eq:VelSolved}) of the main text.

\setcounter{subsection}{6}
\subsection{Section \thesubsection: Experimental determination of the surface capacity $C\rms$}
\label{subsec:determinationCs}

\begin{figure}[H]
    \centering
    \includegraphics[width=0.8\columnwidth]{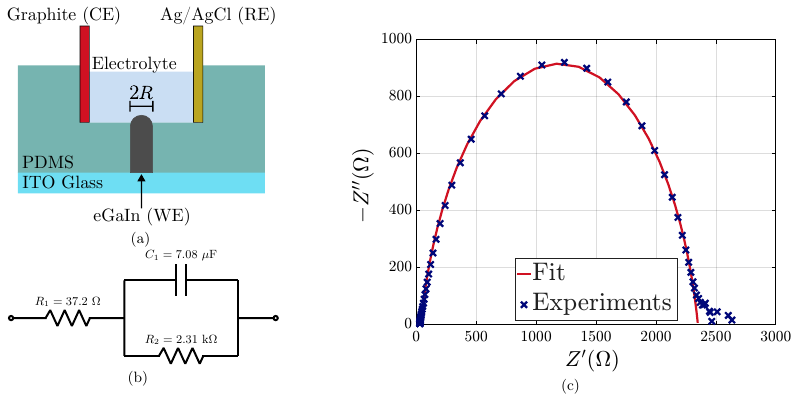}
    \caption{\label{fig:eisUT}(a) Sketch of the experimental setup, (b) equivalent circuit fit and (c) measured results: Nyquist Plot showing the negative imaginary part of the impedance, $-Z''$ versus the real one \cite{lazanasElecImpe23}.}
\end{figure}

To experimentally measure the capacitance of the electrical double layer of EGaIn, we performed electrochemical impedance spectroscopy (EIS) in 1 M NaOH using an Metrohm electrochemical workstation (Autolab PGSTAT302N) with Nova 2.1.3 software. Figure \ref{fig:eisUT}a shows the device setup, where EGaIn was injected into a 4 mm-diameter hole to contact the ITO glass substrate before adding the electrolyte, to ensure that there is no electrolyte between the glass and the metal. Impedance spectra were recorded at the open-circuit potential (OCP = –1.452 V vs. Ag/AgCl (3 M KCl)) using a 10 mV AC perturbation over a frequency range of 10$^5$ to 0.01 Hz. 

The Nyquist plot (Figure \ref{fig:eisUT}c) shows a semicircle without 45$^\circ$ straight line at low frequencies, indicating that there is no mass transport hence the interfacial dynamics of our system is not limited. Consequently, we fitted the Nyquist plot with a simple circuit (\ref{fig:eisUT}b); capturing both the solution resistance ($R_1$) and the semicircle due to interfacial charging~\cite{lazanasElecImpe23}. Thus, performing the fitting in Nova, the capacitance was determined to be \SI{7.08}{\micro\farad} (following Hsu-Mansfeld equation \cite{lazanasElecImpe23}). Considering that the droplet surface can be reasonably approximated as a half-sphere, the surface capacitance was calculated as $\Tilde{C}_\rms = C_1/(2 \pi R^2)$ = \SI{28.18}{\micro\farad\per\centi\meter\squared}. Kramers–Kronig validation confirmed the reliability of the EIS data, with $\chi^2$ values on the order of 10$^{-5}$. Furthermore, this measurement is directly comparable to what obtained by Khan \emph{et al.} in \cite{khanGianSwit14} where the value shown for $C_\rms$, in the absence of surface oxidation, is \SI{30}{\micro\farad\per\centi\meter\squared}.

\setlength{\extrarowheight}{3cm}

\setcounter{subsection}{7}
\subsection{Section \thesubsection: Extreme elongation of a drop for a large continuous excitation voltage}

\begin{figure}[H]
    \centering
    \includegraphics[width=0.6\textwidth]{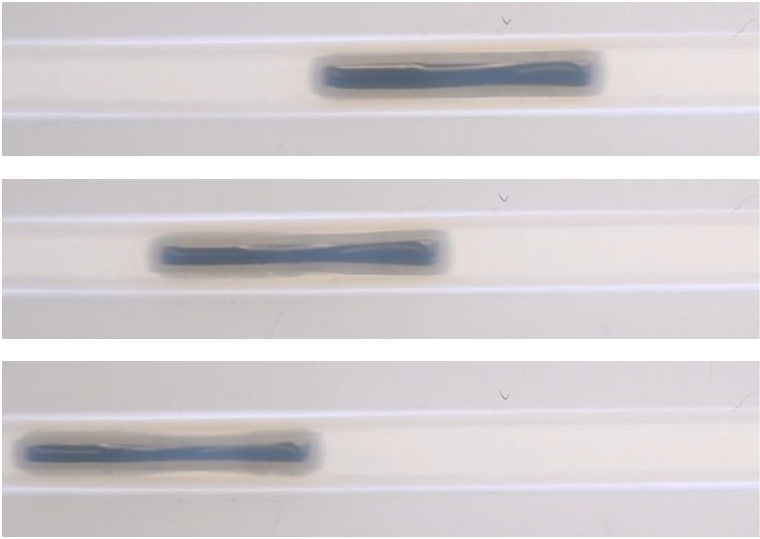}
    \caption{\label{fig:elongated_drop}Snapshots of a video in which a drop of EGaIn clearly elongates inside the capillary as it translates to the left. This video corresponds to imposing a constant voltage $E_0 =$ \SI{5}{\volt} in a capillary of radius $R =$ \SI{0.79}{\milli\meter}. We can clearly see the formation of a liquid electrolyte film around the waist of the metal drop.}
\end{figure}

Since the liquid metal and electrolyte are incompressible, any elongation of the drop is accompanied by a redistribution of the surrounding electrolyte film while preserving the drop volume. Consequently, the assumption of constant drop length employed in the model is well justified over most of the parameter range investigated. To prove these assumptions, we performed an experiment with high applied voltage of 5 V (Fig. \ref{fig:elongated_drop}). Under these conditions, we risk irreversible electrochemical side reactions (the very reason why we kept voltages low in the experiments described in the main text), but under these conditions we can observe stretching and corresponding increase in thickness of the electrolyte thin layer that separates the droplet from the walls (as has been described in ref 18). These observations prove that in our experiments the thin film is present justifying the use of our model.

\setcounter{subsection}{8}
\subsection{Section \thesubsection: Alternative mechanisms that could lead to a resonance behavior}

For completeness, we include here an analysis of two other mechanisms that could explain the existence of the maximum velocity displayed by the liquid metal drop based on possible resonances.\\

\textbf{A bubble trapped in the tube:} Although we do not see any bubble in the tube in our experiments, we find illustrative to compute the size that a bubble should have to exhibit a resonance frequency comparable to the ones observed in our experiments, about $f_\mathrm{res} \sim 1$ Hz. To that end, we use the so-called Minnaert's frequency \cite{leighton2012acoustic}:
\begin{equation}
f_\mathrm{res} = \frac{1}{2\pi} \sqrt{\frac{3\gamma P_0}{\rho_l R_0^2}},
\end{equation}
which gives the resonance frequency of a gas bubble of radius $R_0$ when immersed in a liquid of density $\rho_l$ at ambient pressure $P_0$ and subject to pressure oscillations. Here, $\gamma$ is the specific heat ratio of the gas in the bubble. Assuming the tube entraps an air ($\gamma = 1.4$) bubble at ambient pressure ($P_0 = 10^5$ Pa, as the tube is open at its ends), the radius that such a bubble must have to resonate at a frequency around $f_\mathrm{res} = 1$ Hz would be $R_0 = 3.26$ m. That is, the required bubble size is way too large for the peak frequencies we observe.\\

\textbf{Resonance due to the tube's compliance:} A flexible tube can indeed sustain a resonant behavior if forced at a frequency close to the inverse of the travelling time of pressure waves inside. Since our tube is open at both ends, the travelling time would be $\tau = L_\rmt / 2 c$, where $c$ is the wave speed of inertio-elastic waves due to the tube walls compliance. This velocity can be computed using standard equations taken from the literature \cite{fung2013biomechanics}:
\begin{equation}
c = \frac{E e}{2 \rho_l R},
\end{equation}
where $E$ is the Young's modulus of the tube's material, $\rho_l$ the density of the liquid that fills the tube, $e$ the wall thickness and $R$ the inner radius. We take the following values for our tube: $e = 0.8$ mm, $R = 0.8$ mm, $\rho_l = 10^3$ kg/m$^3$ and $E = 575$ MPa (Designerdata, \emph{Polytetrafluoroethylene (Teflon)}, \url{https://designerdata.nl/materials/plastics/thermo-plastics/polytetrafluoroethylene-(teflon)}. Last visited: September 20$^\mathrm{th}$, 2026)., which yields $c = 534$ m/s and, finally, a traveling time $\tau \approx 0.1$ ms. This corresponds to a frequency about 10,000 times larger than the one we observe in our experiments. This allows us to also rule out the possibility that the frequency peak we observed is caused by a tube's mechanical resonance. Last but not least, as we described in the manuscript, the capillary tube is encased in a rigid acrylic bloc, so the frequency that we estimate is still well below what should be observable in the experiment.


\begin{thebibliography}{31}%
\makeatletter
\providecommand \@ifxundefined [1]{%
 \@ifx{#1\undefined}
}%
\providecommand \@ifnum [1]{%
 \ifnum #1\expandafter \@firstoftwo
 \else \expandafter \@secondoftwo
 \fi
}%
\providecommand \@ifx [1]{%
 \ifx #1\expandafter \@firstoftwo
 \else \expandafter \@secondoftwo
 \fi
}%
\providecommand \natexlab [1]{#1}%
\providecommand \enquote  [1]{``#1''}%
\providecommand \bibnamefont  [1]{#1}%
\providecommand \bibfnamefont [1]{#1}%
\providecommand \citenamefont [1]{#1}%
\providecommand \href@noop [0]{\@secondoftwo}%
\providecommand \href [0]{\begingroup \@sanitize@url \@href}%
\providecommand \@href[1]{\@@startlink{#1}\@@href}%
\providecommand \@@href[1]{\endgroup#1\@@endlink}%
\providecommand \@sanitize@url [0]{\catcode `\\12\catcode `\$12\catcode
  `\&12\catcode `\#12\catcode `\^12\catcode `\_12\catcode `\%12\relax}%
\providecommand \@@startlink[1]{}%
\providecommand \@@endlink[0]{}%
\providecommand \url  [0]{\begingroup\@sanitize@url \@url }%
\providecommand \@url [1]{\endgroup\@href {#1}{\urlprefix }}%
\providecommand \urlprefix  [0]{URL }%
\providecommand \Eprint [0]{\href }%
\providecommand \doibase [0]{https://doi.org/}%
\providecommand \selectlanguage [0]{\@gobble}%
\providecommand \bibinfo  [0]{\@secondoftwo}%
\providecommand \bibfield  [0]{\@secondoftwo}%
\providecommand \translation [1]{[#1]}%
\providecommand \BibitemOpen [0]{}%
\providecommand \bibitemStop [0]{}%
\providecommand \bibitemNoStop [0]{.\EOS\space}%
\providecommand \EOS [0]{\spacefactor3000\relax}%
\providecommand \BibitemShut  [1]{\csname bibitem#1\endcsname}%
\let\auto@bib@innerbib\@empty
\bibitem [{\citenamefont {Dickey}\ \emph {et~al.}(2008)\citenamefont {Dickey},
  \citenamefont {Chiechi}, \citenamefont {Larsen}, \citenamefont {Weiss},
  \citenamefont {Weitz},\ and\ \citenamefont {Whitesides}}]{dickeyEuteGall08}%
  \BibitemOpen
  \bibfield  {author} {\bibinfo {author} {\bibfnamefont {M.~D.}\ \bibnamefont
  {Dickey}}, \bibinfo {author} {\bibfnamefont {R.~C.}\ \bibnamefont {Chiechi}},
  \bibinfo {author} {\bibfnamefont {R.~J.}\ \bibnamefont {Larsen}}, \bibinfo
  {author} {\bibfnamefont {E.~A.}\ \bibnamefont {Weiss}}, \bibinfo {author}
  {\bibfnamefont {D.~A.}\ \bibnamefont {Weitz}},\ and\ \bibinfo {author}
  {\bibfnamefont {G.~M.}\ \bibnamefont {Whitesides}},\ }\bibfield  {title}
  {\bibinfo {title} {Eutectic {{Gallium-Indium}} ({{EGaIn}}): {{A Liquid Metal
  Alloy}} for the {{Formation}} of {{Stable Structures}} in {{Microchannels}}
  at {{Room Temperature}}},\ }\href@noop {} {\bibfield  {journal} {\bibinfo
  {journal} {Advanced Functional Materials}\ }\textbf {\bibinfo {volume}
  {18}},\ \bibinfo {pages} {1097} (\bibinfo {year} {2008})}\BibitemShut
  {NoStop}%
\bibitem [{\citenamefont {Dickey}(2021)}]{dickeyLiquMeta21}%
  \BibitemOpen
  \bibfield  {author} {\bibinfo {author} {\bibfnamefont {M.~D.}\ \bibnamefont
  {Dickey}},\ }\bibfield  {title} {\bibinfo {title} {Liquid metals at room
  temperature},\ }\href@noop {} {\bibfield  {journal} {\bibinfo  {journal}
  {Physics Today}\ }\textbf {\bibinfo {volume} {74}},\ \bibinfo {pages} {30}
  (\bibinfo {year} {2021})}\BibitemShut {NoStop}%
\bibitem [{\citenamefont {Khan}\ \emph
  {et~al.}(2014{\natexlab{a}})\citenamefont {Khan}, \citenamefont {Eaker},
  \citenamefont {Bowden},\ and\ \citenamefont {Dickey}}]{khanGianSwit14}%
  \BibitemOpen
  \bibfield  {author} {\bibinfo {author} {\bibfnamefont {M.~R.}\ \bibnamefont
  {Khan}}, \bibinfo {author} {\bibfnamefont {C.~B.}\ \bibnamefont {Eaker}},
  \bibinfo {author} {\bibfnamefont {E.~F.}\ \bibnamefont {Bowden}},\ and\
  \bibinfo {author} {\bibfnamefont {M.~D.}\ \bibnamefont {Dickey}},\ }\bibfield
   {title} {\bibinfo {title} {Giant and switchable surface activity of liquid
  metal via surface oxidation},\ }\href@noop {} {\bibfield  {journal} {\bibinfo
   {journal} {Proc. Natl. Acad. Sci.}\ }\textbf {\bibinfo {volume} {111}},\
  \bibinfo {pages} {14047} (\bibinfo {year} {2014}{\natexlab{a}})}\BibitemShut
  {NoStop}%
\bibitem [{\citenamefont {Eaker}\ and\ \citenamefont
  {Dickey}(2016)}]{eakerLiquMeta16}%
  \BibitemOpen
  \bibfield  {author} {\bibinfo {author} {\bibfnamefont {C.~B.}\ \bibnamefont
  {Eaker}}\ and\ \bibinfo {author} {\bibfnamefont {M.~D.}\ \bibnamefont
  {Dickey}},\ }\bibfield  {title} {\bibinfo {title} {Liquid metal actuation by
  electrical control of interfacial tension},\ }\href@noop {} {\bibfield
  {journal} {\bibinfo  {journal} {Applied Physics Reviews}\ }\textbf {\bibinfo
  {volume} {3}},\ \bibinfo {pages} {031103} (\bibinfo {year}
  {2016})}\BibitemShut {NoStop}%
\bibitem [{\citenamefont {Eaker}\ \emph {et~al.}(2017)\citenamefont {Eaker},
  \citenamefont {Hight}, \citenamefont {O'Regan}, \citenamefont {Dickey},\ and\
  \citenamefont {Daniels}}]{eakerOxidFing17}%
  \BibitemOpen
  \bibfield  {author} {\bibinfo {author} {\bibfnamefont {C.~B.}\ \bibnamefont
  {Eaker}}, \bibinfo {author} {\bibfnamefont {D.~C.}\ \bibnamefont {Hight}},
  \bibinfo {author} {\bibfnamefont {J.~D.}\ \bibnamefont {O'Regan}}, \bibinfo
  {author} {\bibfnamefont {M.~D.}\ \bibnamefont {Dickey}},\ and\ \bibinfo
  {author} {\bibfnamefont {K.~E.}\ \bibnamefont {Daniels}},\ }\bibfield
  {title} {\bibinfo {title} {Oxidation-{{Mediated Fingering}} in {{Liquid
  Metals}}},\ }\href@noop {} {\bibfield  {journal} {\bibinfo  {journal}
  {Physical Review Letters}\ }\textbf {\bibinfo {volume} {119}},\ \bibinfo
  {pages} {174502} (\bibinfo {year} {2017})}\BibitemShut {NoStop}%
\bibitem [{\citenamefont {Beni}\ \emph {et~al.}(1982)\citenamefont {Beni},
  \citenamefont {Hackwood},\ and\ \citenamefont {Jackel}}]{beniContElec82}%
  \BibitemOpen
  \bibfield  {author} {\bibinfo {author} {\bibfnamefont {G.}~\bibnamefont
  {Beni}}, \bibinfo {author} {\bibfnamefont {S.}~\bibnamefont {Hackwood}},\
  and\ \bibinfo {author} {\bibfnamefont {J.~L.}\ \bibnamefont {Jackel}},\
  }\bibfield  {title} {\bibinfo {title} {Continuous electrowetting effect},\
  }\href@noop {} {\bibfield  {journal} {\bibinfo  {journal} {Applied Physics
  Letters}\ }\textbf {\bibinfo {volume} {40}},\ \bibinfo {pages} {912}
  (\bibinfo {year} {1982})}\BibitemShut {NoStop}%
\bibitem [{\citenamefont {Gough}\ \emph {et~al.}(2014)\citenamefont {Gough},
  \citenamefont {Morishita}, \citenamefont {Dang}, \citenamefont {Hu},
  \citenamefont {Shiroma},\ and\ \citenamefont {Ohta}}]{goughContElec14}%
  \BibitemOpen
  \bibfield  {author} {\bibinfo {author} {\bibfnamefont {R.~C.}\ \bibnamefont
  {Gough}}, \bibinfo {author} {\bibfnamefont {A.~M.}\ \bibnamefont
  {Morishita}}, \bibinfo {author} {\bibfnamefont {J.~H.}\ \bibnamefont {Dang}},
  \bibinfo {author} {\bibfnamefont {W.}~\bibnamefont {Hu}}, \bibinfo {author}
  {\bibfnamefont {W.~A.}\ \bibnamefont {Shiroma}},\ and\ \bibinfo {author}
  {\bibfnamefont {A.~T.}\ \bibnamefont {Ohta}},\ }\bibfield  {title} {\bibinfo
  {title} {Continuous {{Electrowetting}} of {{Non-toxic Liquid Metal}} for {{RF
  Applications}}},\ }\href@noop {} {\bibfield  {journal} {\bibinfo  {journal}
  {IEEE Access}\ }\textbf {\bibinfo {volume} {2}},\ \bibinfo {pages} {874}
  (\bibinfo {year} {2014})}\BibitemShut {NoStop}%
\bibitem [{\citenamefont {Wang}\ \emph {et~al.}(2015)\citenamefont {Wang},
  \citenamefont {Trlica}, \citenamefont {Khan}, \citenamefont {Dickey},\ and\
  \citenamefont {Adams}}]{wangRecoLiqu15}%
  \BibitemOpen
  \bibfield  {author} {\bibinfo {author} {\bibfnamefont {M.}~\bibnamefont
  {Wang}}, \bibinfo {author} {\bibfnamefont {C.}~\bibnamefont {Trlica}},
  \bibinfo {author} {\bibfnamefont {M.~R.}\ \bibnamefont {Khan}}, \bibinfo
  {author} {\bibfnamefont {M.~D.}\ \bibnamefont {Dickey}},\ and\ \bibinfo
  {author} {\bibfnamefont {J.~J.}\ \bibnamefont {Adams}},\ }\bibfield  {title}
  {\bibinfo {title} {A reconfigurable liquid metal antenna driven by
  electrochemically controlled capillarity},\ }\href@noop {} {\bibfield
  {journal} {\bibinfo  {journal} {Journal of Applied Physics}\ }\textbf
  {\bibinfo {volume} {117}},\ \bibinfo {pages} {194901} (\bibinfo {year}
  {2015})}\BibitemShut {NoStop}%
\bibitem [{\citenamefont {Otero~Mart{\'i}nez}\ \emph
  {et~al.}(2022)\citenamefont {Otero~Mart{\'i}nez}, \citenamefont
  {Rodr{\'i}guez~Rodr{\'i}guez}, \citenamefont {Shen}, \citenamefont {Tong},
  \citenamefont {Wong},\ and\ \citenamefont
  {Garc{\'i}a~Armada}}]{oteromaLiquReco22}%
  \BibitemOpen
  \bibfield  {author} {\bibinfo {author} {\bibfnamefont {J.}~\bibnamefont
  {Otero~Mart{\'i}nez}}, \bibinfo {author} {\bibfnamefont {J.}~\bibnamefont
  {Rodr{\'i}guez~Rodr{\'i}guez}}, \bibinfo {author} {\bibfnamefont
  {Y.}~\bibnamefont {Shen}}, \bibinfo {author} {\bibfnamefont {K.-F.}\
  \bibnamefont {Tong}}, \bibinfo {author} {\bibfnamefont {K.-K.}\ \bibnamefont
  {Wong}},\ and\ \bibinfo {author} {\bibfnamefont {A.}~\bibnamefont
  {Garc{\'i}a~Armada}},\ }\bibfield  {title} {\bibinfo {title} {Toward {{Liquid
  Reconfigurable Antenna Arrays}} for {{Wireless Communications}}},\
  }\href@noop {} {\bibfield  {journal} {\bibinfo  {journal} {IEEE
  Communications Magazine}\ }\textbf {\bibinfo {volume} {60}},\ \bibinfo
  {pages} {145} (\bibinfo {year} {2022})}\BibitemShut {NoStop}%
\bibitem [{\citenamefont {Jackel}\ \emph {et~al.}(1983)\citenamefont {Jackel},
  \citenamefont {Hackwood}, \citenamefont {Veselka},\ and\ \citenamefont
  {Beni}}]{jackelElecSwit83}%
  \BibitemOpen
  \bibfield  {author} {\bibinfo {author} {\bibfnamefont {J.~L.}\ \bibnamefont
  {Jackel}}, \bibinfo {author} {\bibfnamefont {S.}~\bibnamefont {Hackwood}},
  \bibinfo {author} {\bibfnamefont {J.~J.}\ \bibnamefont {Veselka}},\ and\
  \bibinfo {author} {\bibfnamefont {G.}~\bibnamefont {Beni}},\ }\bibfield
  {title} {\bibinfo {title} {Electrowetting switch for multimode optical
  fibers},\ }\href@noop {} {\bibfield  {journal} {\bibinfo  {journal} {Applied
  Optics}\ }\textbf {\bibinfo {volume} {22}},\ \bibinfo {pages} {1765}
  (\bibinfo {year} {1983})}\BibitemShut {NoStop}%
\bibitem [{\citenamefont {Zhao}\ \emph {et~al.}(2023)\citenamefont {Zhao},
  \citenamefont {Soni}, \citenamefont {Lee}, \citenamefont {Nijhuis},\ and\
  \citenamefont {Xiang}}]{zhaoSmarEute23}%
  \BibitemOpen
  \bibfield  {author} {\bibinfo {author} {\bibfnamefont {Z.}~\bibnamefont
  {Zhao}}, \bibinfo {author} {\bibfnamefont {S.}~\bibnamefont {Soni}}, \bibinfo
  {author} {\bibfnamefont {T.}~\bibnamefont {Lee}}, \bibinfo {author}
  {\bibfnamefont {C.~A.}\ \bibnamefont {Nijhuis}},\ and\ \bibinfo {author}
  {\bibfnamefont {D.}~\bibnamefont {Xiang}},\ }\bibfield  {title} {\bibinfo
  {title} {Smart {{Eutectic Gallium}}--{{Indium}}: {{From Properties}} to
  {{Applications}}},\ }\href@noop {} {\bibfield  {journal} {\bibinfo  {journal}
  {Advanced Materials}\ }\textbf {\bibinfo {volume} {35}},\ \bibinfo {pages}
  {2203391} (\bibinfo {year} {2023})}\BibitemShut {NoStop}%
\bibitem [{\citenamefont {Lee}\ and\ \citenamefont
  {Kim}(2000)}]{leeSurfTens00}%
  \BibitemOpen
  \bibfield  {author} {\bibinfo {author} {\bibfnamefont {J.}~\bibnamefont
  {Lee}}\ and\ \bibinfo {author} {\bibfnamefont {C.-J.}\ \bibnamefont {Kim}},\
  }\bibfield  {title} {\bibinfo {title} {Surface tension driven microactuation
  based on continuous electrowetting},\ }\href@noop {} {\bibfield  {journal}
  {\bibinfo  {journal} {Journal of Microelectromechanical Systems}\ }\textbf
  {\bibinfo {volume} {9}},\ \bibinfo {pages} {171} (\bibinfo {year}
  {2000})}\BibitemShut {NoStop}%
\bibitem [{\citenamefont {Tang}\ \emph {et~al.}(2015)\citenamefont {Tang},
  \citenamefont {Lin}, \citenamefont {Joshipura}, \citenamefont {Khoshmanesh},\
  and\ \citenamefont {Dickey}}]{tangSteeLiqu15}%
  \BibitemOpen
  \bibfield  {author} {\bibinfo {author} {\bibfnamefont {S.-Y.}\ \bibnamefont
  {Tang}}, \bibinfo {author} {\bibfnamefont {Y.}~\bibnamefont {Lin}}, \bibinfo
  {author} {\bibfnamefont {I.~D.}\ \bibnamefont {Joshipura}}, \bibinfo {author}
  {\bibfnamefont {K.}~\bibnamefont {Khoshmanesh}},\ and\ \bibinfo {author}
  {\bibfnamefont {M.~D.}\ \bibnamefont {Dickey}},\ }\bibfield  {title}
  {\bibinfo {title} {Steering liquid metal flow in microchannels using low
  voltages},\ }\href@noop {} {\bibfield  {journal} {\bibinfo  {journal} {Lab on
  a Chip}\ }\textbf {\bibinfo {volume} {15}},\ \bibinfo {pages} {3905}
  (\bibinfo {year} {2015})}\BibitemShut {NoStop}%
\bibitem [{sup()}]{suppmat}%
  \BibitemOpen
  \href@noop {} {}\bibinfo {note} {See Supplemental Material at [URL will be
  inserted by publisher] for details on the experimental methodology,
  theoretical model derivations, and additional discussion of non-dominant
  physical effects. The Suppl. Mat. includes Refs.~\cite{buchnerDielRela99,
  xuEffeOxid12, lu2007diffuse, dataZenodo, leighton2012acoustic,
  fung2013biomechanics}}\BibitemShut {NoStop}%
\bibitem [{\citenamefont {Qu{\'e}r{\'e}}(1997)}]{quereInerCapi97}%
  \BibitemOpen
  \bibfield  {author} {\bibinfo {author} {\bibfnamefont {D.}~\bibnamefont
  {Qu{\'e}r{\'e}}},\ }\bibfield  {title} {\bibinfo {title} {Inertial
  capillarity},\ }\href@noop {} {\bibfield  {journal} {\bibinfo  {journal}
  {Europhysics Letters (EPL)}\ }\textbf {\bibinfo {volume} {39}},\ \bibinfo
  {pages} {533} (\bibinfo {year} {1997})}\BibitemShut {NoStop}%
\bibitem [{\citenamefont {Landau}(1984)}]{landauElecCont84}%
  \BibitemOpen
  \bibfield  {author} {\bibinfo {author} {\bibfnamefont {L.~D.}\ \bibnamefont
  {Landau}},\ }\href@noop {} {\emph {\bibinfo {title} {Electrodynamics of
  Continuous Media}}},\ \bibinfo {edition} {2nd}\ ed.,\ \bibinfo {series}
  {Course of Theoretical Physics}\ No.~\bibinfo {number} {8}\ (\bibinfo
  {publisher} {Pergamon Press},\ \bibinfo {address} {Oxford [etc},\ \bibinfo
  {year} {1984})\BibitemShut {NoStop}%
\bibitem [{\citenamefont {Kirby}(2010)}]{kirbyMicrNano10}%
  \BibitemOpen
  \bibfield  {author} {\bibinfo {author} {\bibfnamefont {B.~J.}\ \bibnamefont
  {Kirby}},\ }\href@noop {} {\emph {\bibinfo {title} {Micro- and {{Nanoscale
  Fluid Mechanics}}: {{Transport}} in {{Microfluidic Devices}}}}}\ (\bibinfo
  {publisher} {Cambridge University Press},\ \bibinfo {address} {Cornaell
  University, NY},\ \bibinfo {year} {2010})\BibitemShut {NoStop}%
\bibitem [{\citenamefont {Balestra}\ \emph {et~al.}(2018)\citenamefont
  {Balestra}, \citenamefont {Zhu},\ and\ \citenamefont
  {Gallaire}}]{balestraViscTayl18}%
  \BibitemOpen
  \bibfield  {author} {\bibinfo {author} {\bibfnamefont {G.}~\bibnamefont
  {Balestra}}, \bibinfo {author} {\bibfnamefont {L.}~\bibnamefont {Zhu}},\ and\
  \bibinfo {author} {\bibfnamefont {F.}~\bibnamefont {Gallaire}},\ }\bibfield
  {title} {\bibinfo {title} {Viscous {{Taylor}} droplets in axisymmetric and
  planar tubes: From {{Bretherton}}'s theory to empirical models},\ }\href@noop
  {} {\bibfield  {journal} {\bibinfo  {journal} {Microfluidics and
  Nanofluidics}\ }\textbf {\bibinfo {volume} {22}},\ \bibinfo {pages} {67}
  (\bibinfo {year} {2018})}\BibitemShut {NoStop}%
\bibitem [{\citenamefont {Khan}\ \emph
  {et~al.}(2014{\natexlab{b}})\citenamefont {Khan}, \citenamefont {Trlica},
  \citenamefont {So}, \citenamefont {Valeri},\ and\ \citenamefont
  {Dickey}}]{khanInflWate14}%
  \BibitemOpen
  \bibfield  {author} {\bibinfo {author} {\bibfnamefont {M.~R.}\ \bibnamefont
  {Khan}}, \bibinfo {author} {\bibfnamefont {C.}~\bibnamefont {Trlica}},
  \bibinfo {author} {\bibfnamefont {J.-H.}\ \bibnamefont {So}}, \bibinfo
  {author} {\bibfnamefont {M.}~\bibnamefont {Valeri}},\ and\ \bibinfo {author}
  {\bibfnamefont {M.~D.}\ \bibnamefont {Dickey}},\ }\bibfield  {title}
  {\bibinfo {title} {Influence of {{Water}} on the {{Interfacial Behavior}} of
  {{Gallium Liquid Metal Alloys}}},\ }\href@noop {} {\bibfield  {journal}
  {\bibinfo  {journal} {ACS Applied Materials \& Interfaces}\ }\textbf
  {\bibinfo {volume} {6}},\ \bibinfo {pages} {22467} (\bibinfo {year}
  {2014}{\natexlab{b}})}\BibitemShut {NoStop}%
\bibitem [{\citenamefont {Womersley}(1955)}]{womersleMethCalc55}%
  \BibitemOpen
  \bibfield  {author} {\bibinfo {author} {\bibfnamefont {J.~R.}\ \bibnamefont
  {Womersley}},\ }\bibfield  {title} {\bibinfo {title} {Method for the
  calculation of velocity, rate of flow and viscous drag in arteries when the
  pressure gradient is known},\ }\href@noop {} {\bibfield  {journal} {\bibinfo
  {journal} {The Journal of Physiology}\ }\textbf {\bibinfo {volume} {127}},\
  \bibinfo {pages} {553} (\bibinfo {year} {1955})}\BibitemShut {NoStop}%
\bibitem [{\citenamefont {Lazanas}\ and\ \citenamefont
  {Prodromidis}(2023)}]{lazanasElecImpe23}%
  \BibitemOpen
  \bibfield  {author} {\bibinfo {author} {\bibfnamefont {A.~C.}\ \bibnamefont
  {Lazanas}}\ and\ \bibinfo {author} {\bibfnamefont {M.~I.}\ \bibnamefont
  {Prodromidis}},\ }\bibfield  {title} {\bibinfo {title} {Electrochemical
  {{Impedance Spectroscopy}}},\ }\href@noop {} {\bibfield  {journal} {\bibinfo
  {journal} {ACS Measurement Science Au}\ }\textbf {\bibinfo {volume} {3}},\
  \bibinfo {pages} {162} (\bibinfo {year} {2023})}\BibitemShut {NoStop}%
\bibitem [{\citenamefont {Dollet}\ \emph {et~al.}(2020)\citenamefont {Dollet},
  \citenamefont {Lorenceau},\ and\ \citenamefont
  {Gallaire}}]{dolletTranExpo20}%
  \BibitemOpen
  \bibfield  {author} {\bibinfo {author} {\bibfnamefont {B.}~\bibnamefont
  {Dollet}}, \bibinfo {author} {\bibfnamefont {{\'E}.}~\bibnamefont
  {Lorenceau}},\ and\ \bibinfo {author} {\bibfnamefont {F.}~\bibnamefont
  {Gallaire}},\ }\bibfield  {title} {\bibinfo {title} {Transition from
  {{Exponentially Damped}} to {{Finite-Time Arrest Liquid Oscillations
  Induced}} by {{Contact Line Hysteresis}}},\ }\href@noop {} {\bibfield
  {journal} {\bibinfo  {journal} {Physical Review Letters}\ }\textbf {\bibinfo
  {volume} {124}},\ \bibinfo {pages} {104502} (\bibinfo {year}
  {2020})}\BibitemShut {NoStop}%
\bibitem [{\citenamefont {{Ruiz-Guti{\'e}rrez}}\ \emph
  {et~al.}(2022)\citenamefont {{Ruiz-Guti{\'e}rrez}}, \citenamefont
  {Armstrong}, \citenamefont {L{\'e}v{\^e}que}, \citenamefont {Michel},
  \citenamefont {Pagonabarraga}, \citenamefont {Wells}, \citenamefont
  {{Hern{\'a}ndez-Machado}},\ and\ \citenamefont
  {{Ledesma-Aguilar}}}]{ruiz-gutLongCros22}%
  \BibitemOpen
  \bibfield  {author} {\bibinfo {author} {\bibfnamefont {{\'E}.}~\bibnamefont
  {{Ruiz-Guti{\'e}rrez}}}, \bibinfo {author} {\bibfnamefont {S.}~\bibnamefont
  {Armstrong}}, \bibinfo {author} {\bibfnamefont {S.}~\bibnamefont
  {L{\'e}v{\^e}que}}, \bibinfo {author} {\bibfnamefont {C.}~\bibnamefont
  {Michel}}, \bibinfo {author} {\bibfnamefont {I.}~\bibnamefont
  {Pagonabarraga}}, \bibinfo {author} {\bibfnamefont {G.~G.}\ \bibnamefont
  {Wells}}, \bibinfo {author} {\bibfnamefont {A.}~\bibnamefont
  {{Hern{\'a}ndez-Machado}}},\ and\ \bibinfo {author} {\bibfnamefont
  {R.}~\bibnamefont {{Ledesma-Aguilar}}},\ }\bibfield  {title} {\bibinfo
  {title} {The long cross-over dynamics of capillary imbibition},\ }\href@noop
  {} {\bibfield  {journal} {\bibinfo  {journal} {Journal of Fluid Mechanics}\
  }\textbf {\bibinfo {volume} {939}},\ \bibinfo {pages} {A39} (\bibinfo {year}
  {2022})}\BibitemShut {NoStop}%
\bibitem [{\citenamefont {Schnitzer}\ \emph {et~al.}(2013)\citenamefont
  {Schnitzer}, \citenamefont {Frankel},\ and\ \citenamefont
  {Yariv}}]{schnitzer2013electrokinetic}%
  \BibitemOpen
  \bibfield  {author} {\bibinfo {author} {\bibfnamefont {O.}~\bibnamefont
  {Schnitzer}}, \bibinfo {author} {\bibfnamefont {I.}~\bibnamefont {Frankel}},\
  and\ \bibinfo {author} {\bibfnamefont {E.}~\bibnamefont {Yariv}},\ }\bibfield
   {title} {\bibinfo {title} {Electrokinetic flows about conducting drops},\
  }\href@noop {} {\bibfield  {journal} {\bibinfo  {journal} {Journal of Fluid
  Mechanics}\ }\textbf {\bibinfo {volume} {722}},\ \bibinfo {pages} {394}
  (\bibinfo {year} {2013})}\BibitemShut {NoStop}%
\bibitem [{\citenamefont {Babatain}\ \emph {et~al.}(2025)\citenamefont
  {Babatain}, \citenamefont {Park}, \citenamefont {Harraz}, \citenamefont
  {Kilic~Afsar}, \citenamefont {Honnet}, \citenamefont {Lov}, \citenamefont
  {Labrune}, \citenamefont {Dickey},\ and\ \citenamefont
  {Ishii}}]{babatain2025programmable}%
  \BibitemOpen
  \bibfield  {author} {\bibinfo {author} {\bibfnamefont {W.}~\bibnamefont
  {Babatain}}, \bibinfo {author} {\bibfnamefont {C.}~\bibnamefont {Park}},
  \bibinfo {author} {\bibfnamefont {D.~M.}\ \bibnamefont {Harraz}}, \bibinfo
  {author} {\bibfnamefont {O.}~\bibnamefont {Kilic~Afsar}}, \bibinfo {author}
  {\bibfnamefont {C.}~\bibnamefont {Honnet}}, \bibinfo {author} {\bibfnamefont
  {S.}~\bibnamefont {Lov}}, \bibinfo {author} {\bibfnamefont {J.-B.}\
  \bibnamefont {Labrune}}, \bibinfo {author} {\bibfnamefont {M.~D.}\
  \bibnamefont {Dickey}},\ and\ \bibinfo {author} {\bibfnamefont
  {H.}~\bibnamefont {Ishii}},\ }\bibfield  {title} {\bibinfo {title}
  {Programmable continuous electrowetting of liquid metal for reconfigurable
  electronics},\ }\href@noop {} {\bibfield  {journal} {\bibinfo  {journal}
  {Advanced Materials}\ ,\ \bibinfo {pages} {e06383}} (\bibinfo {year}
  {2025})}\BibitemShut {NoStop}%
\bibitem [{\citenamefont {Buchner}\ \emph {et~al.}(1999)\citenamefont
  {Buchner}, \citenamefont {Hefter}, \citenamefont {May},\ and\ \citenamefont
  {Sipos}}]{buchnerDielRela99}%
  \BibitemOpen
  \bibfield  {author} {\bibinfo {author} {\bibfnamefont {R.}~\bibnamefont
  {Buchner}}, \bibinfo {author} {\bibfnamefont {G.}~\bibnamefont {Hefter}},
  \bibinfo {author} {\bibfnamefont {P.~M.}\ \bibnamefont {May}},\ and\ \bibinfo
  {author} {\bibfnamefont {P.}~\bibnamefont {Sipos}},\ }\bibfield  {title}
  {\bibinfo {title} {Dielectric {{Relaxation}} of {{Dilute Aqueous NaOH}},
  {{NaAl}}({{OH}})4, and {{NaB}}({{OH}})4},\ }\href@noop {} {\bibfield
  {journal} {\bibinfo  {journal} {Journal of Physical Chemistry B}\ }\textbf
  {\bibinfo {volume} {103}},\ \bibinfo {pages} {11186} (\bibinfo {year}
  {1999})}\BibitemShut {NoStop}%
\bibitem [{\citenamefont {Xu}\ \emph {et~al.}(2012)\citenamefont {Xu},
  \citenamefont {Oudalov}, \citenamefont {Guo}, \citenamefont {Jaeger},\ and\
  \citenamefont {Brown}}]{xuEffeOxid12}%
  \BibitemOpen
  \bibfield  {author} {\bibinfo {author} {\bibfnamefont {Q.}~\bibnamefont
  {Xu}}, \bibinfo {author} {\bibfnamefont {N.}~\bibnamefont {Oudalov}},
  \bibinfo {author} {\bibfnamefont {Q.}~\bibnamefont {Guo}}, \bibinfo {author}
  {\bibfnamefont {H.~M.}\ \bibnamefont {Jaeger}},\ and\ \bibinfo {author}
  {\bibfnamefont {E.}~\bibnamefont {Brown}},\ }\bibfield  {title} {\bibinfo
  {title} {Effect of oxidation on the mechanical properties of liquid gallium
  and eutectic gallium-indium},\ }\href@noop {} {\bibfield  {journal} {\bibinfo
   {journal} {Physics of Fluids}\ }\textbf {\bibinfo {volume} {24}},\ \bibinfo
  {pages} {063101} (\bibinfo {year} {2012})}\BibitemShut {NoStop}%
\bibitem [{\citenamefont {Lu}\ \emph {et~al.}(2007)\citenamefont {Lu},
  \citenamefont {Glasner}, \citenamefont {Bertozzi},\ and\ \citenamefont
  {Kim}}]{lu2007diffuse}%
  \BibitemOpen
  \bibfield  {author} {\bibinfo {author} {\bibfnamefont {H.-W.}\ \bibnamefont
  {Lu}}, \bibinfo {author} {\bibfnamefont {K.}~\bibnamefont {Glasner}},
  \bibinfo {author} {\bibfnamefont {A.}~\bibnamefont {Bertozzi}},\ and\
  \bibinfo {author} {\bibfnamefont {C.-J.}\ \bibnamefont {Kim}},\ }\bibfield
  {title} {\bibinfo {title} {A diffuse-interface model for electrowetting drops
  in a hele-shaw cell},\ }\href@noop {} {\bibfield  {journal} {\bibinfo
  {journal} {Journal of Fluid Mechanics}\ }\textbf {\bibinfo {volume} {590}},\
  \bibinfo {pages} {411} (\bibinfo {year} {2007})}\BibitemShut {NoStop}%
\bibitem [{\citenamefont {Otero~Martinez}\ and\ \citenamefont
  {Rodríguez-Rodríguez}(2026)}]{dataZenodo}%
  \BibitemOpen
  \bibfield  {author} {\bibinfo {author} {\bibfnamefont {J.}~\bibnamefont
  {Otero~Martinez}}\ and\ \bibinfo {author} {\bibfnamefont {J.}~\bibnamefont
  {Rodríguez-Rodríguez}},\ }\bibfield  {title} {\bibinfo {title} {Transient
  response of drops of egain subject to {CEW} to be applied to {FA}},\ }\href
  {https://doi.org/10.5281/zenodo.23343043} {10.5281/zenodo.23343043} (\bibinfo
  {year} {2026})\BibitemShut {NoStop}%
\bibitem [{\citenamefont {Leighton}(2012)}]{leighton2012acoustic}%
  \BibitemOpen
  \bibfield  {author} {\bibinfo {author} {\bibfnamefont {T.}~\bibnamefont
  {Leighton}},\ }\href@noop {} {\emph {\bibinfo {title} {The acoustic
  bubble}}}\ (\bibinfo  {publisher} {Academic press},\ \bibinfo {year}
  {2012})\BibitemShut {NoStop}%
\bibitem [{fun(2013)}]{fung2013biomechanics}%
  \BibitemOpen
  \href@noop {} {\emph {\bibinfo {title} {Biomechanics: circulation}}}\
  (\bibinfo  {publisher} {Springer Science \& Business Media},\ \bibinfo {year}
  {2013})\BibitemShut {NoStop}%
\end{thebibliography}
\end{document}